\documentclass[aps,pre,floatfix,superscriptaddress,preprint,10pt]{revtex4-1}

\usepackage{bm}
\usepackage{empheq}
\usepackage{float}
\usepackage{amsfonts}
\usepackage{amssymb}
\usepackage{amsmath}
\usepackage[capitalise,noabbrev]{cleveref}
\usepackage[titletoc]{appendix}
\usepackage{multirow}
\usepackage{physics}
\usepackage{graphicx}
\usepackage[a4paper]{geometry}

\everymath={\displaystyle}
\crefname{equation}{}{}
\crefrangelabelformat{equation}{(#3#1#4)--(#5\crefstripprefix{#1}{#2}#6)}

\newcommand{\etal}{{\it et al.\ }}
\newcommand{\ie}{i.e.\ }
\newcommand{\eg}{e.g.\ }
\newcommand{\Oh}{O}

\newcommand{\neff}{n_{\rm eff}}
\newcommand{\RIne}{n_{\rm e}}
\newcommand{\RIno}{n_{\rm o}}

\newcommand{\thetaNS}{\theta_{\rm NS}}
\newcommand{\thetaGN}{\theta_{\rm GN}}
\newcommand{\thetaNShat}{\hat{\theta}_{\rm NS}}
\newcommand{\thetaGNhat}{\hat{\theta}_{\rm GN}}
\newcommand{\n}{\bm{n}}

\newcommand{\CNS}{C_{\rm NS}}
\newcommand{\CGN}{C_{\rm GN}}
\newcommand{\CNShat}{\hat{C}_{\rm NS}}
\newcommand{\CGNhat}{\hat{C}_{\rm GN}}
\newcommand{\gammaNS}{\gamma_{\rm NS}}

\newcommand{\hmax}{h_{\rm m}}
\newcommand{\hiso}{h_{\rm I}}
\newcommand{\betaL}{\beta^{-}}
\newcommand{\betaR}{\beta^{+}}

\newcommand{\hc}{h_{\rm c}}
\newcommand{\xc}{x_{\rm c}}

\newcommand{\pI}{{p_0}_{\rm I}}
\newcommand{\tpO}{\tilde{p}_0}

\newcommand{\DelE}{\Delta E}
\newcommand{\DelEP}{\Delta E_{\mathcal{P}}}
\newcommand{\DelEH}{\Delta E_{\mathcal{H}}}
\newcommand{\Etot}{E_{\rm tot}}
\newcommand{\Esurf}{E_{\rm surf}}
\newcommand{\Eelas}{E_{\rm elas}}
\newcommand{\EB}{E_{\rm NS}}
\newcommand{\ET}{E_{\rm GN}}

\newcommand{\unitx}{\bm{\hat{x}}}
\newcommand{\unity}{\bm{\hat{y}}}
\newcommand{\unitz}{\bm{\hat{z}}}

\newcommand{\Psol}{\mathcal{P}}
\newcommand{\Hsol}{\mathcal{H}}
\newcommand{\Dsol}{\mathcal{D}}
\newcommand{\DPsol}{\mathcal{D}_\mathcal{P}}
\newcommand{\DHsol}{\mathcal{D}_\mathcal{H}}

\newcommand{\tK}{\tilde{K}}
\newcommand{\tA}{\tilde{A}}
\newcommand{\tL}{\tilde{L}}
\newcommand{\tl}{\tilde{l}}
\newcommand{\tCNS}{\tilde{C}_{\rm NS}}
\newcommand{\tCGN}{\tilde{C}_{\rm GN}}
\newcommand{\tgammaNS}{\tilde{\gamma}_{\rm NS}}
\newcommand{\tgammaGN}{\tilde{\gamma}_{\rm GN}}
\newcommand{\td}{\tilde{d}}
\newcommand{\tx}{\tilde{x}}
\newcommand{\ty}{\tilde{y}}
\newcommand{\tz}{\tilde{z}}
\newcommand{\thmax}{\tilde{h}_{\rm m}}
\newcommand{\thiso}{\tilde{h}_{\rm I}}
\newcommand{\tilh}{\tilde{h}}
\newcommand{\tbetaL}{\tilde{\beta}^{-}}
\newcommand{\tbetaR}{\tilde{\beta}^{+}}
\newcommand{\tbetaLR}{\tilde{\beta}^{\pm}}
\newcommand{\tgrad}{\tilde{\grad}}
\newcommand{\tEtot}{\tilde{E}_{\rm tot}}
\newcommand{\twns}{\tilde{\omega}_{\rm NS}}
\newcommand{\twgn}{\tilde{\omega}_{\rm GN}}
\newcommand{\tW}{\tilde{W}}

\begin{document}

\title{Weak-anchoring effects in a thin pinned ridge of nematic liquid crystal}
\author{Joseph R.\ L.\ \surname{Cousins}}
\email{joseph.cousins@strath.ac.uk, joseph.cousins@glasgow.ac.uk}
\affiliation{Department of Mathematics and Statistics, University of Strathclyde, Livingstone Tower, 26 Richmond Street, Glasgow G1 1XH, United Kingdom}
\affiliation{School of Mathematics and Statistics, University of Glasgow, University Place, Glasgow G12 8QQ, United Kingdom}
\author{Akhshay S.\ \surname{Bhadwal}}
\email{akhshay.bhadwal@ntu.ac.uk}
\affiliation{SOFT Group, School of Science and Technology, Nottingham Trent University, Clifton Lane, Nottingham NG11 8NS, United Kingdom}
\author{Lindsey T.\ \surname{Corson}}
\email{lindsey.corson@strath.ac.uk}
\affiliation{Department of Mathematics and Statistics, University of Strathclyde, Livingstone Tower, 26 Richmond Street, Glasgow G1 1XH, United Kingdom}
\author{Brian R.\ \surname{Duffy}}
\email{b.r.duffy@strath.ac.uk}
\affiliation{Department of Mathematics and Statistics, University of Strathclyde, Livingstone Tower, 26 Richmond Street, Glasgow G1 1XH, United Kingdom}
\author{Ian C.\ \surname{Sage}}
\email{ian.sage@ntu.ac.uk}
\affiliation{SOFT Group, School of Science and Technology, Nottingham Trent University, Clifton Lane, Nottingham NG11 8NS, United Kingdom}
\author{Carl V.\ \surname{Brown}}
\email{carl.brown@ntu.ac.uk}
\affiliation{SOFT Group, School of Science and Technology, Nottingham Trent University, Clifton Lane, Nottingham NG11 8NS, United Kingdom}
\author{Nigel J.\ \surname{Mottram}}
\email{nigel.mottram@glasgow.ac.uk}
\affiliation{School of Mathematics and Statistics, University of Glasgow, University Place, Glasgow G12 8QQ, United Kingdom}
\author{Stephen K.\ \surname{Wilson}}
\thanks{Author for correspondence}
\email{s.k.wilson@strath.ac.uk}
\affiliation{Department of Mathematics and Statistics, University of Strathclyde, Livingstone Tower, 26 Richmond Street, Glasgow G1 1XH, United Kingdom}

\date{7th July 2022, revised 21st December 2022}

\begin{abstract}
A theoretical investigation of weak-anchoring effects in a thin two-dimensional pinned static ridge of nematic liquid crystal resting on a flat solid substrate in an atmosphere of passive gas is performed.
Specifically, we solve a reduced version of the general system of governing equations recently derived by Cousins \etal [{\it Proc.\ Roy.\ Soc.\ A}, {\bf 478}(2259):20210849, 2022] valid for a symmetric thin ridge under the one-constant approximation of the Frank--Oseen bulk elastic energy with pinned contact lines to determine the shape of the ridge and the behaviour of the director within it.
Numerical investigations covering a wide range of parameter values indicate that the energetically-preferred solutions can be classified in terms of the Jenkins--Barratt--Barbero--Barberi critical thickness into five qualitatively different types of solution.
In particular, the theoretical results suggest that anchoring breaking occurs close to the contact lines.
The theoretical predictions are supported by the results of physical experiments for a ridge of the nematic 4'-pentyl-4-biphenylcarbonitrile (5CB).
In particular, these experiments show that the homeotropic anchoring at the gas--nematic interface is broken close to the contact lines by the stronger rubbed planar anchoring at the nematic--substrate interface.
A comparison between the experimental values of and the theoretical predictions for the effective refractive index of the ridge gives a first estimate of the anchoring strength of an interface between air and 5CB to be $(9.80\pm1.12)\times10^{-6}\,{\rm N m}^{-1}$ at a temperature of $(22\pm1.5)^\circ$C.
\end{abstract}

\maketitle

\section{Introduction}
\label{sec:introduction}

\subsection{Background}
\label{sec:background}

Since the late 1960s, technological interest in nematic liquid crystals (nematics) has largely been focused on their use in Liquid Crystal Displays (LCDs) \cite{patent:Heilmeier1968, Jones2017, Jones2018}.
However, more recently, applications of nematics have taken advantage of their intrinsically viscoelastic nature, which has led to the development of novel microfluidic devices with applications in particle transport \cite{Cavallaro2011,Lavrentovich2014,Bhadwal2020}, molecule sensing \cite{Lin2011}, and medicine \cite{Sengupta2013}.
In addition, their optoelectrical properties make them well suited for new adaptive-lens technologies \cite{LensFab2017,LensReview2019}, microelectronic components \cite{Gentili2012,Zou2018}, and diffraction gratings \cite{Brown2009,Blow2013}.
Many of these emerging technologies are complicated multiphase systems that involve interfaces between the nematic, a solid substrate, and an atmosphere of gas (or another isotropic fluid).

Much of the recent theoretical interest in situations that involve nematic interfaces has focused on spontaneous dewetting transitions, such as spinodal dewetting and nucleation events, in which a thin film of nematic ruptures and forms droplets or ridges \cite{Demirel1999,Vandenbrouck1999,Braun2000,vanderWielen2000,Ziherl2000,Ostovskii2001,vanEffenterre2003,LinCummings2013,Dhara2020}.
These spontaneous dewetting transitions have been studied with a range of theoretical approaches, from continuum theory \cite{LinCummings2013,LamCummings2018,LamCummings2020} to statistical mechanics \cite{Vanzo2016,Rull2017,Allen2019}, that often fail to account accurately for the nematic molecular alignment forces on the nematic interfaces (commonly called \emph{anchoring}).
An accurate characterisation of anchoring on the nematic interfaces is particularly important in situations that involve a three-phase contact line at which the nematic--substrate interface, the gas--nematic interface, and the gas--substrate interface meet.
In particular, assuming that there are infinitely strong anchoring forces on both of the nematic interfaces will, in general, lead to inconsistent predictions for the nematic molecular orientation at the contact line.
One way to avoid such difficulties is to assume that the nematic molecular orientation approaches the appropriate uniform state dictated by the (typically stronger) anchoring force on the nematic--substrate interface as the contact line is approached (see, for example, Lam \etal \cite{LamCummings2018,LamCummings2020}).
Another way is to assume the presence of a thin precursor film on the substrate, which has the effect of removing the contact line entirely (see, for example, Lin \etal \cite{LinCummings2013B}).
In what follows, we instead allow the competition between the anchoring forces on the nematic interfaces to determine the molecular orientation close to the contact line.

In the present work, we perform a theoretical investigation of weak-anchoring effects in a thin two-dimensional pinned static ridge of nematic resting on a flat solid substrate in an atmosphere of passive gas.
The ridge has a nematic--substrate interface and a gas--nematic interface (\ie the nematic free surface), and pinned nematic--substrate--gas three-phase contact lines (\ie the contact lines are at fixed locations on the substrate) at its left-hand and right-hand edges.
As well as being of fundamental scientific interest in their own right, pinned ridges of liquid crystal occur in
the self-organisation of columnar discotic liquid crystals into ridges \cite{Bramble2010,Zou2018,Bramble2021},
nematic diffraction gratings \cite{Brown2009,Blow2013}, and
ridges of nematic mixtures \cite{Bao2021}.
The behaviour of a two-dimensional static ridge of nematic which is not necessarily thin and may have either pinned or unpinned contact lines is described by the general system of governing equations (including both nematic Young and nematic Young--Laplace equations) recently derived by Cousins \etal \cite{Cousins2022}.
However, these authors did not attempt to solve the full system of equations, but instead used the nematic Young equations they derived to determine the continuous and discontinuous transitions that occur between the equilibrium states of complete wetting, partial wetting, and complete dewetting.
In the present work, we take a somewhat different approach and solve a reduced version of the general system of governing equations valid for a symmetric thin ridge under the one-constant approximation of the Frank--Oseen bulk elastic energy with pinned contact lines to determine the shape of the ridge and the behaviour of the director within it.
The theoretical predictions are supported by the results of physical experiments for a ridge of the nematic 4'-pentyl-4-biphenylcarbonitrile (5CB).
The results obtained give insight into the behaviour of nematic systems with interfaces and contact lines, and, in particular, give a first estimate of the anchoring strength of an interface between air and 5CB.

\subsection{Nematic anchoring}
\label{sec:anchoring}

In the present work we assume that the bulk energy of the nematic ridge is described by the Frank--Oseen elastic energy density, which depends on the average nematic molecular orientation, represented by a unit vector $\n$ called the \emph{director}, and its spatial gradients \cite{ISBOOK2004}.
We also assume that the energy densities of the nematic--substrate and the gas--nematic interfaces are described by the Rapini--Papoular interface energy density \cite{RapiniPapoularWA1969}, which describes the energetic preference of the nematic molecular alignment forces on the interfaces to align the director, and is commonly called \emph{weak anchoring}.

We consider two types of weak anchoring, namely \emph{homeotropic weak anchoring}, where there is an energetic preference for the director to align normally to the interface, and \emph{homogeneous planar weak anchoring} (henceforth referred to as simply \emph{planar weak anchoring}), where there is an energetic preference for the director to align in a particular preferred tangential direction to the interface.
Planar weak anchoring at a nematic--substrate interface is often called \emph{rubbed planar anchoring}, and the preferred tangential direction is often called the \emph{rubbing direction} \cite{ISBOOK2004}.
Planar weak anchoring with no preferred tangential direction is known as \emph{planar degenerate anchoring}.
In reality,
the director may prefer to align at a small angle to a normal or a tangential alignment,
a phenomenon commonly known as pretilt \cite{SONINBOOK1995}.
Since a close-to-zero pretilt is obtainable in physical experiments \cite{Cui2012},
we assume that there is no pretilt in the present work.
The strength of the energetic preference for a particular alignment of the director on an interface is measured by a parameter called the \emph{anchoring strength}.
In the limit in which the anchoring strength becomes infinitely large, the energetic preference for homeotropic or planar alignment fixes the director on the interface to align exactly normally or tangentially, respectively, to it.
This situation is often called \emph{strong anchoring} or, more accurately, \emph{infinite anchoring}.
When the anchoring strength is zero there is no energetic preference for homeotropic or planar alignment on the interface, and the director on the interface is determined by bulk forces.
This situation is called \emph{zero anchoring}.
In the present work, we will consider
weak zenithal anchoring (\ie allow in-plane rotation of the director towards or away from the interface normal) but
infinite azimuthal anchoring (\ie not allow rotation of the director around the interface normal),
so that the director remains in plane.

Systems with a gas--nematic interface can involve opposing anchoring preferences on the nematic--substrate and the gas--nematic interfaces \cite{SONINBOOK1995}, \eg when homeotropic alignment is preferred on the gas--nematic interface and planar alignment is preferred on the nematic--substrate interface, or vice versa.
This situation is called \emph{antagonistic anchoring}, while \emph{non-antagonistic anchoring} refers to the opposite situation in which the same alignment is preferred on the two interfaces.
Note that, since the preferred alignments are measured \emph{relative} to the interfaces, which are, in general, not parallel, even situations with non-antagonistic anchoring can involve competing anchoring forces.

\subsection{Nematic contact lines}
\label{sec:contactlines}

As mentioned in \cref{sec:background}, careful analysis of the anchoring forces at the nematic interfaces is fundamental to understanding the behaviour of the director close to a contact line.
Two special scenarios for the behaviour of the director close to a contact line were considered by Rey \cite{Rey2003,Rey2007}, namely infinite planar anchoring on both interfaces and equal (finite) planar weak anchoring on the two interfaces.
In these studies, Rey highlighted the possibility of \emph{anchoring breaking} occurring close to the contact line.
Anchoring breaking occurs when the torque on the director at an interface due to the weak anchoring on that interface is overcome by the torque due to other effects (in the present case, by the torque due to the weak anchoring on the other interface), and the resulting director orientation is significantly different from that preferred by the anchoring.
Cousins \etal \cite{Cousins2022} assumed that anchoring breaking occurs close to the contact line and used the nematic Young equations they derived to determine the continuous and discontinuous transitions between the equilibrium states of complete wetting, partial wetting, and complete dewetting that can occur.
In the present work we instead allow the behaviour of the director to be determined by the competition between the anchoring forces on the nematic interfaces, and find that anchoring breaking does indeed occur close to the contact lines, validating the analysis of \cite{Cousins2022} in the present situation.
Rey \cite{Rey2003,Rey2007} also discussed the possibility of the occurrence of defects, or disclination lines in the two-dimensional situation we consider, at the contact line.
At disclination lines, a description of the nematic only in terms of the director is no longer valid, and large elastic distortions give rise to a local increase in the elastic energy density and a reduction in the orientational order about the director \cite{SchopohlSluckin1987}.
In the present work, we assume that anchoring breaking is always energetically preferred to the formation of a disclination line at the contact line.
This assumption is valid provided that the nematic coherence length \cite{Mottram2010} is smaller than the surface extrapolation length \cite{SONINBOOK1995} of either the nematic--substrate interface or the gas--nematic interface (see \cref{sec:governing} for details).

\subsection{The Jenkins--Barratt--Barbero--Barberi critical thickness}
\label{sec:hc}

Of particular importance to the present work is a pioneering but perhaps overlooked theoretical contribution published in 1974 by Jenkins and Barratt \cite{JenkinsBarratt1974}, who, after detailing the theoretical description of a gas--nematic interface and a nematic contact line, investigated a static layer of nematic of uniform thickness with antagonistic anchoring due to weak anchoring on the gas--nematic interface and infinite anchoring on the nematic--substrate interface.
They showed that when the bulk elastic energy density for the nematic layer takes the Frank--Oseen form, there is a critical thickness of the layer below which a uniform director field (oriented according to the preferred alignment on the interface with the stronger anchoring, \ie the nematic--substrate interface) is energetically preferred, and above which a distorted director field is energetically preferred.
In 1983, Barbero and Barberi \cite{BarberoandBarberi1983} independently obtained the corresponding result for antagonistic anchoring due to weak anchoring on both interfaces.
The critical thickness of the layer is often referred to as the Barbero--Barberi critical thickness \cite{BarberoandBarberi1983}, but, to give due credit to the work of Jenkins and Barratt \cite{JenkinsBarratt1974}, we refer to it as \emph{the Jenkins--Barratt--Barbero--Barberi critical thickness} or, for brevity, simply \emph{the critical thickness} \cite{JenkinsBarratt1974,BarberoandBarberi1983}.
This critical thickness is of particular importance for a ridge of nematic because the height of the ridge is, by definition, zero at the contact lines, but may exceed the critical thickness elsewhere.
Understanding the behaviour of the director in any regions of the ridge in which its height is greater than this critical thickness, and hence a distorted director field is energetically favourable, and regions of the ridge in which the height is less than the critical thickness, and hence a uniform director field is energetically favourable, is a key aspect of the present work.

\subsection{Outline of the present work}
\label{sec:outline}

In \cref{sec:ridge} we derive a reduced version of the general system of governing equations for a two-dimensional static ridge of nematic valid for a symmetric thin ridge under the one-constant approximation of the Frank--Oseen bulk elastic energy with pinned contact lines.
We obtain analytical solutions for situations with a uniform director field in \cref{sec:uniform} and numerical solutions for situations with a distorted director field in \cref{sec:distorted}, showing that five qualitatively different types of solution occur.
In \cref{sec:preferred} we determine the regions of parameter space in which the different types of solution introduced in \cref{sec:uniform,sec:distorted} are energetically preferred.
Finally, in \cref{sec:experiments} the theoretical predictions are compared with the results of physical experiments for a ridge of 5CB.

\section{Model formulation}
\label{sec:ridge}

\subsection{Geometry of the problem}

\begin{figure}[tp]
\centering
\includegraphics[width=0.75\linewidth]{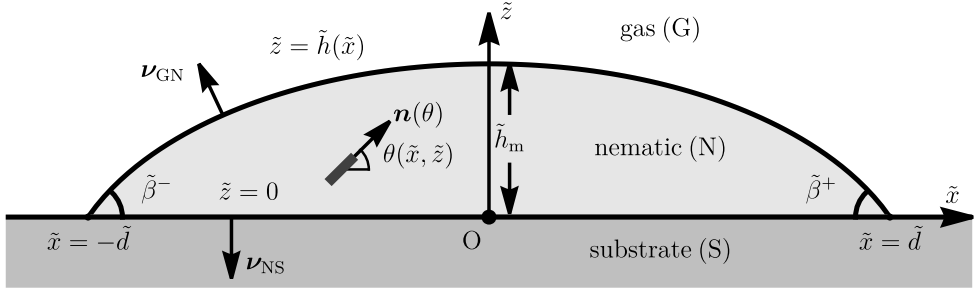}
\caption{
A schematic of a two-dimensional pinned static ridge of nematic (N) with prescribed cross-sectional area $\tA$ resting on a flat solid substrate (S) in an atmosphere of passive gas (G), bounded by a gas--nematic interface at $\tz=\tilh(\tx)$ and a nematic--substrate interface at $\tz=0$, with pinned contact lines at $\tx=\pm \td$.
The Cartesian coordinates $\tx$ and $\tz$ (with the $\ty$-axis out of the page), the contact angles $\tbetaL$ and $\tbetaR$, and the outward unit normals of the nematic--substrate interface and gas--nematic interface $\bm{\nu}_{\rm NS}$ and $\bm{\nu}_{\rm GN}$, respectively, are indicated.
A typical director $\n$ with director angle $\theta(\tilde{x},\tilde{z})$ and the height at the middle of the ridge $\thmax$ are also shown.
}
\label{fig:1}
\end{figure}

We consider a two-dimensional pinned static ridge of nematic (N) with prescribed cross-sectional area $\tA$ resting on a flat solid substrate (S) in an atmosphere of passive gas (G), as shown in \cref{fig:1}.
The nematic ridge is bounded by the gas--nematic interface at $\tz=\tilh(\tx)$,
where $\tilh$ is the height of the ridge, and
the nematic--substrate interface at $\tz=0$,
where $\tx$, $\ty$, and $\tz$ are the Cartesian coordinates indicated in \cref{fig:1}.
The nematic--substrate and gas--nematic interfaces meet at pinned contact lines at $\tx=\pm\td$,
where $\td$ is the prescribed constant semi-width of the ridge, so that
\begin{equation}\label{EqBC}
\tilh=0 \quad \mbox{at} \quad \tx=\pm\td.
\end{equation}
As the nematic ridge is two dimensional,
the height of the ridge $\tilh$ and the positions of the pinned contact lines do not vary in the $\ty$-direction,
and so the contact lines form two infinitely-long parallel lines in the $\ty$-direction.
The unknown height at the middle of the ridge is denoted $\thmax$, \ie $\thmax=\tilh(0)$.
The unknown contact angles formed between the nematic--substrate and gas--nematic interfaces are denoted by $\tbetaL$ and $\tbetaR$,
and satisfy $\tan\tbetaLR=\mp\tilh_{\tx}$ at $\tx=\pm\td$,
where the subscript denotes differentiation.
Following Cousins \etal \cite{Cousins2022},
we assume that the director $\n$ is confined to the $(\tx,\tz)$-plane,
and thus can be expressed in terms of the angle between the director and the $\tx$-axis,
denoted by $\theta=\theta(\tx,\tz)$, as $\n=\n(\theta)$ given by
\begin{align}\label{n2D}
\n=\cos\theta\,\unitx+\sin\theta\,\unitz,
\end{align}
where $\unitx$ and $\unitz$ are unit vectors in the $\tx$- and $\tz$-direction, respectively.
Note that
while the cross-sectional area $\tA$ and semi-width $\td$ of the ridge are prescribed,
the shape of the gas--nematic interface $\tilh$
(and, in particular, the height at the middle of the ridge $\thmax$ and the contact angles $\tbetaL$ and $\tbetaR$) and
the director field $\n$
are determined as part of the solution to the problem.

\subsection{Governing equations for a pinned nematic ridge}
\label{sec:governing}

The general system of governing equations and boundary conditions for a static ridge of nematic were derived by Cousins \etal \cite{Cousins2022} by minimisation of the free energy, which includes contributions from the Frank--Oseen bulk elastic energy density $\tW=\tW(\theta,\theta_{\tx},\theta_{\tz})$ \cite{ISBOOK2004}, gravitational potential energy density $\tilde{\psi}_{\rm g}=\tilde{\psi}_{\rm g}(\tx,\tz)$, and the Rapini--Papoular interface energy densities for the nematic--substrate interface $\twns=\twns(\theta)$ and the gas--nematic interface $\twgn=\twgn(\theta,\tilh_{\tx})$ \cite{RapiniPapoularWA1969}, subject to an area constraint.

In the present work, unlike in Cousins \etal \cite{Cousins2022}, we make the one-constant approximation to the Frank--Oseen bulk elastic energy density (obtained by setting ${\tK}_1={\tK}_3=\tK$ in \cite{Cousins2022}) to obtain a mathematically tractable system of equations in which $\tW$ is given by
\begin{equation}\label{OFenergy}
\tW = \dfrac{\tK}{2} \left[(\tgrad \cdot \n)^2 + (\tgrad \times \n)^2 + \tgrad \cdot \left((\n \cdot \tgrad)\n - (\tgrad \cdot \n)\n \right)\right]
= \dfrac{\tK}{2} \left( \theta_{\tx}^2 +\theta_{\tz}^2 \right),
\end{equation}
where $\tgrad=\unitx \partial/\partial \tx+\unity\partial/\partial \ty+\unitz\partial/\partial\tz$.
Following Cousins \etal \cite{Cousins2022}, we take the Rapini--Papoular interface energy densities $\twns$ and $\twgn$ to be
\begin{align}
\twns &=\tgammaNS +\dfrac{\tCNS}{4} \left( 1 - 2 (\n \cdot \bm{\nu}_{\rm NS})^2 \right)
= \tgammaNS+\dfrac{\tCNS}{4} \cos 2\theta, \label{RPenergy1} \\
\twgn &=\tgammaGN +\dfrac{\tCGN}{4} \left( 1 - 2 (\n \cdot \bm{\nu}_{\rm GN})^2 \right)
= \tgammaGN+\dfrac{\tCGN}{4(1+\tilh_{\tx}^2)} \left[(1-\tilh_{\tx}^2) \cos 2\theta+2\tilh_{\tx} \sin 2\theta \right], \label{RPenergy2}
\end{align}
respectively, where $\bm{\nu}_{\rm NS}$ and $\bm{\nu}_{\rm GN}$ are the outward unit normals of the nematic--substrate and gas--nematic interfaces, respectively, as shown in \cref{fig:1}.
%
%
The constant material parameters $\tK\,(>0)$, $\tgammaNS\,(>0)$, $\tgammaGN\,(>0)$, $\tCNS$, and $\tCGN$ appearing in \cref{OFenergy,RPenergy1,RPenergy2} are the one-constant elastic constant, the interfacial tension of the nematic--substrate interface, the interfacial tension of the gas--nematic interface, the anchoring strength of the nematic--substrate interface, and the anchoring strength of the gas--nematic interface, respectively.
If $\tilde{C}_i>0$ then a director normal to the interface (\ie a homeotropic alignment) is preferred on the nematic--substrate interface ($i={\rm NS}$) or gas--nematic interface ($i={\rm GN}$), whereas if $\tilde{C}_i<0$ then a director tangent to the interface (\ie a planar alignment) is preferred.
Experimental techniques for the measurement of $\tCNS$ are well established, and hence values of $\tCNS$ are readily available \cite{AnchoringMeasurement1985,Yokoyama1988,AnchoringMeasurement1999}; however, measurements of $\tCGN$ are more difficult, and hence values of $\tCGN$ are less readily available \cite{SONINBOOK1995}.
Specifically, the anchoring strength of a planar or homeotropic nematic--substrate interface has been measured for a variety of nematic materials and substrates in the range $\vert\tCNS\vert= 10^{-6}$--$10^{-3}\,{\rm N m}^{-1}$ \cite{AnchoringMeasurement1985,Yokoyama1988,SONINBOOK1995,AnchoringMeasurement1999}, whereas, to the best of our knowledge, the only experimental measurements of the anchoring strength of a gas--nematic interface are $\tCGN>10^{-5}\,{\rm N m}^{-1}$ for the interface between air and the nematic mixture ZLI 2860 \cite{FreeSurfaceAnchoring1997}, and $\tCGN>4\times10^{-4}\,{\rm N m}^{-1}$ for the interface between air and the nematic p-methoxy-benzylidene-p-n-butyl aniline (MBBA) \cite{Chiarelli1983}.
Given these anchoring strengths and a typical one-constant elastic constant of $\tK=10^{-11}$ \cite{PHYSICALPROPERTIES2001}, we find that the surface extrapolation lengths of the nematic--substrate and the gas--nematic interfaces, which are given by $\tK/\vert\tCNS\vert$ and $\tK/\vert\tCGN\vert$, respectively, are in the range $10\,$nm--$10\,\mu$m.
The nematic coherence length of typical nematic materials is $4\,$nm \cite{Khayyatzadeh2015} for a wide range of temperatures away from the nematic--isotropic transition temperature \cite{Mottram2010}, and we therefore expect that our assumption that anchoring breaking is always energetically preferred to the formation of a disclination line at the contact line is valid, except possibly close to the nematic--isotropic transition temperature.
Gas--nematic interfaces are often assumed to exhibit homeotropic weak anchoring \cite{SONINBOOK1995}. We note, however, that the experimental studies of Feng \etal \cite{Feng2012} and Bao \etal \cite{Bao2021} indicate that planar degenerate anchoring can also occur on such interfaces.
We therefore allow the gas--nematic interface to have either homeotropic ($\tCGN>0$) or planar ($\tCGN<0$) weak anchoring.

The isotropic components of the Rapini--Papoular interface energy densities \cref{RPenergy1,RPenergy2} (\ie the interfacial tensions $\tgammaNS$ and $\tgammaGN$, also commonly known as surface tensions) are more readily measured than the anchoring strengths.
For instance, the interfacial tension between air and the nematic 5CB has been measured as $\tgammaGN=4.0\times10^{-2}\,{\rm N m}^{-1}$ \cite{Dhara2020}, and the interfacial tension between the solid poly(methyl methacrylate) (PMMA) and 5CB has been measured as $\tgammaNS=4.051\times10^{-2}\,{\rm N m}^{-1}$ \cite{Dhara2020}.

In the present work we assume that gravitational effects can be neglected compared to surface-tension effects, and therefore we require that the semi-width of the ridge $\td$ is much smaller than the capillary length $\tl=(\tgammaGN/(\tilde{\rho}\tilde{g}))^{1/2}$, where $\tilde{\rho}$ is the constant density of the nematic and $\tilde{g}$ is the magnitude of acceleration due to gravity \cite{deGennesWetting1985}.
This assumption is appropriate for the physical experiments presented in \cref{sec:experiments}, where $\tl=2\times10^{-3}\,$m and $\td=6\times10^{-4}\,$m, and for many other applications,
including
ridges of discotic liquid crystals used in new semi-conductor applications \cite{Zou2018},
nematic diffraction gratings \cite{Brown2009,Blow2013},
ridges of nematic mixtures \cite{Bao2021}, and possibly for
nematic droplets used in LCD manufacturing \cite{Cousins2019,Cousins2020}.

In the absence of gravitational effects, and with the one-constant approximation, the general system of governing equations and boundary conditions derived by Cousins \etal \cite{Cousins2022} consists of the bulk elastic equation given by
\begin{equation}\label{EqBulkF}
0=\theta_{\tx\tx}+\theta_{\tz\tz},
\end{equation}
and three interfacial equations, namely the weak anchoring (\ie balance-of-couple) conditions on the nematic--substrate interface and the gas--nematic interface, and the nematic Young--Laplace equation on the gas--nematic interface, which are given by
\begin{align}
0 &= -\tK \theta_{\tz}-\dfrac{\tCNS}{2} \sin 2\theta \quad \text{on} \quad \tz=0, \label{EqLSF} \\
0 &= \tK \left(\theta_{\tz}-{\tilh}_{\tx} \theta_{\tx}\right)+\dfrac{\tCGN}{2\sqrt{1+\tilh_{\tx}^2}}\left[(\tilh_{\tx}^2-1)\sin 2\theta+2 \tilh_{\tx}\cos 2\theta\right] \quad \text{on}  \quad \tz=\tilh, \label{EqGLF} \\
0 &= \tpO-\dfrac{\tK}{2}\left(\theta_{\tx}^2+\theta_{\tz}^2\right)+\dfrac{\tgammaGN \tilh_{\tx\tx}}{\left(1+\tilh_{\tx}^2\right)^{3/2}}+\dfrac{\tCGN}{4\left(1+\tilh_{\tx}^2\right)^{5/2}} \Bigg[3 \tilh_{\tx\tx}\left\{(\tilh_{\tx}^2-1)\cos 2\theta-2\tilh_{\tx}\sin 2\theta\right\} \label{EqYLF} \nonumber \\
& \quad +\left(1+\tilh_{\tx}^2\right)\left\{4\cos2\theta\left[\theta_{\tx}-\tilh_{\tx}(1+\tilh_{\tx}^2)\theta_{\tz}\right]+2\sin 2\theta \left[(1-\tilh_{\tx}^4)\theta_{\tz}+\tilh_{\tx}(3+\tilh_{\tx}^2) \theta_{\tx}\right] \right\}\Bigg] \quad \text{on} \quad \tz=\tilh,
\end{align}
respectively, where $\tpO$ is a Lagrange multiplier with the dimensions of pressure associated with the area constraint.

The total energy of the pinned ridge, denoted by $\tEtot$, is given by
\begin{align}\label{Etot}
\tEtot & = \int_{-\td}^{\td} \, \int_0^{\tilh} \tW \, \dd \tz \, \dd \tx + \int_{-\td}^{\td}  \twns \, \dd \tx + \int_{-\td}^{\td} \twgn\,\sqrt{1+\tilh_{\tx}^2} \, \dd \tx.
\end{align}

The general system of governing equations and boundary conditions derived by Cousins \etal \cite{Cousins2022} also includes nematic Young equations which describe the balance of stress at unpinned contact lines.
However, since the present work concerns pinned contact lines, these equations are not relevant here.
The corresponding analysis of a thin ridge with unpinned contact lines will be considered in future work.

\subsection{Governing equations for a thin pinned nematic ridge}
\label{sec:thin}

Since in many of the practical applications described in \cref{sec:background} the nematic ridges and droplets are thin, and in order to enable us to make analytical progress, we henceforth consider the situation in which the ridge is thin.
Specifically, adopting the well-known thin-film (or lubrication) approximation,
we define an appropriate small nondimensional aspect ratio $\epsilon$ in terms of the semi-width $\td$ and the cross-sectional area $\tA$ as
\begin{align}\label{ep}
\epsilon &= \dfrac{\tA}{\td^2} \ll 1,
\end{align}
and assume that the variables scale according to
\begin{align}\label{Scale}
\begin{gathered}
\tx = \td \, x, \qquad \tz = \epsilon \td \, z, \qquad \tilh = \epsilon \td \, h,  \qquad \thmax = \epsilon \td \, \hmax, \\
\tbetaL = \epsilon \, \betaL, \qquad \tbetaR = \epsilon \, \betaR, \qquad \tpO = \frac{\epsilon \tgammaGN}{\td} \, p_0, \\
\tEtot = \td \tgammaGN \, \Etot, \qquad \tW = \dfrac{\epsilon \tgammaGN}{\td} \, W, \qquad  \twns = \tgammaGN \, \omega_{\rm NS}, \qquad \twgn = \tgammaGN \, \omega_{\rm GN}, \\
\tK = \epsilon^3 \td \tgammaGN \, K, \qquad \tgammaNS = \tgammaGN \, \gammaNS, \qquad \tCNS = \epsilon^2 \tgammaGN \, \CNS, \qquad \tCGN = \epsilon^2 \tgammaGN \, \CGN,
\end{gathered}
\end{align}
where quantities with a superposed tilde ($\,\tilde{}\,$) are dimensional and quantities without a superposed tilde are nondimensional.
Note that we have nondimensionalised lengths in the $\tx$-direction with $\td$ and lengths in the $\tz$-direction with $\epsilon \td$, and hence the contact angles are scaled with $\epsilon$.
The interfacial tension $\tgammaNS$ and the anchoring strengths $\tCNS$ and $\tCGN$ are nondimensionalised with the interfacial tension $\tgammaGN$.
In order to study the most interesting regime in which surface tension, anchoring, and bulk elasticity effects are all comparable, we have nondimensionalised the elastic constant $\tK$ and anchoring strengths $\tCNS$ and $\tCGN$ such that contributions of surface tension, anchoring, and bulk elasticity appear at leading order in the reduced equations.
Less interesting regimes for which bulk elasticity effects are either much stronger or much weaker than anchoring effects will be discussed in \cref{sec:limits}.

At leading order in the limit $\epsilon \to 0$, the nondimensional form of the bulk elastic equation \cref{EqBulkF} reduces to
\begin{equation}
\theta_{zz} = 0,
\end{equation}
and hence the director angle $\theta$ is given by
\begin{equation}\label{thinbulk}
\theta = \thetaNS + \left(\thetaGN - \thetaNS\right)\frac{z}{h},
\end{equation}
where $\thetaNS=\thetaNS(x)=\theta(x,0)$ and $\thetaGN=\thetaGN(x)=\theta(x,h(x))$ denote the (as yet unknown) values of $\theta$ on the nematic--substrate interface $z=0$ and on the gas--nematic interface $z=h$, respectively.

The nondimensional forms of equations \cref{EqLSF,EqGLF,EqYLF} reduce to
\begin{equation}\label{thinNS}
K\left(\thetaGN - \thetaNS\right) + \CNS h \sin \thetaNS \cos \thetaNS = 0,
\end{equation}
\begin{equation}\label{thinGN}
K\left(\thetaGN - \thetaNS\right) - \CGN h \sin \thetaGN \cos \thetaGN = 0,
\end{equation}
\begin{equation}\label{thinYL}
p_0 + h_{xx} + \frac{K}{2}\left(\frac{\thetaGN - \thetaNS}{h}\right)^2 = 0,
\end{equation}
respectively.
Note that, unlike the full anchoring condition on the gas--nematic interface \cref{EqGLF}, the leading-order anchoring condition \cref{thinGN} depends only on the local height of the ridge $h$ and not on its derivatives.
In particular, as expected in the present thin-film limit, \cref{thinGN} is identical to the anchoring condition on a flat gas--nematic interface (\ie when $h$ is constant); therefore, $\thetaGN$ behaves as if the gas--nematic interface were locally flat.
A consequence of this, combined with the rotational symmetry of the director, is that if $\theta=\hat{\theta}$ is a solution of \cref{thinNS,thinGN,thinYL} then so is $\theta=q\pi\pm\hat{\theta}$ for any integer $q$.
In the leading-order nematic Young--Laplace equation \cref{thinYL}, the leading-order approximation of the curvature of the gas--nematic interface, namely $h_{xx}$, is coupled to a term that depends on both anchoring and bulk elasticity effects, namely $K\left(\thetaGN-\thetaNS\right)^2/(2h^2)$.

The non-dimensional form of the contact-line conditions \cref{EqBC} is
\begin{align}\label{thinBC2}
h = 0 \quad \mbox{at} \quad x = \pm 1,
\end{align}
and hence it follows from the anchoring conditions \cref{thinNS,thinGN} that $\theta=\thetaNS=\thetaGN$ at $x=\pm 1$.

The nondimensional form of the prescribed cross-sectional area constraint is
\begin{align}\label{thinArea2}
1 &= \int_{-1}^{1} h \, \dd x,
\end{align}
and the leading-order contact angles are given by $\betaL=h_x$ at $x=-1$ and $\betaR=-h_x$ at $x=1$.

The nondimensional form of the total energy \cref{Etot} is
\begin{align}
\Etot & = 2\left(1+\gammaNS\right) + \dfrac{1}{2}\epsilon^2 \DelE + \Oh(\epsilon^3),
\end{align}
showing that $\Etot$ takes the constant value of $2(1+\gammaNS)$ at leading order in $\epsilon$, and that variations in $\Etot$ appear at second order in $\epsilon$ via the term $\DelE=\DelE(h,\thetaNS,\thetaGN)$, which can be decomposed as
\begin{align}\label{thinEtot}
\DelE=\Esurf+\Eelas+\EB+\ET,
\end{align}
where
\begin{align}\label{thinEparts}
\begin{gathered}
\Esurf = \int_{-1}^{1} h_x^2 \, \dd x, \qquad
\Eelas = K \int_{-1}^{1} \dfrac{\left(\thetaGN - \thetaNS\right)^2}{h} \, \dd x, \\
\EB = \dfrac{\CNS}{2} \int_{-1}^{1} \cos 2\thetaNS \, \dd x, \qquad
\ET = \dfrac{\CGN}{2} \int_{-1}^{1} \cos 2\thetaGN \, \dd x
\end{gathered}
\end{align}
are the second-order surface energy $\Esurf$, elastic energy $\Eelas$, nematic--substrate interface anchoring energy $\EB$, and gas--nematic interface anchoring energy $\ET$, respectively.
The special case of an isotropic fluid is recovered by setting $K=\CNS=\CGN=0$, so that the second-order energy reduces to $\DelE=\Esurf$.
The second-order surface energy $\Esurf$ is minimised subject to the contact-line conditions \cref{thinBC2} and the area constraint \cref{thinArea2} by the solution for the height and the Lagrange multiplier for an isotropic ridge in the absence of gravity, denoted $\hiso=\hiso(x)$ and $\pI$, respectively, namely
\begin{align}\label{hiso}
\hiso=\dfrac{3}{4}\left(1-x^2\right) \quad \text{and} \quad \pI=\dfrac{3}{2}.
\end{align}
Inspection of \cref{thinEparts} with \cref{hiso} shows that $\Esurf \ge 3/2$, $\Eelas \ge 0$, $-\vert\CNS\vert \le \EB \le \vert\CNS\vert$, and $-\vert\CGN\vert \le \ET \le \vert\CGN\vert$.

In summary, the equations and boundary conditions for a thin pinned ridge given by the system \cref{thinNS,thinGN,thinYL,thinBC2,thinArea2} involve the unknowns $\thetaNS(x)$, $\thetaGN(x)$, $h(x)$, and $p_0$ and the parameters $K$, $\CNS$, and $\CGN$.
Once the solutions for $\thetaNS$, $\thetaGN$, $h$, and $p_0$ have been determined, the contact angles $\beta^+=-h_x(1)$ and $\beta^-=h_x(-1)$, the director angle given by \cref{thinbulk}, and the second-order energy $\DelE$ given by \cref{thinEtot,thinEparts} can be readily determined.

Note that the intrinsic symmetries of the system \cref{thinNS,thinGN,thinYL,thinBC2,thinArea2} mean that if $\thetaNS=\hat{\theta}_{\rm NS}$ and $\thetaGN=\hat{\theta}_{\rm GN}$ are solutions for $\CNS=\CNShat$ and $\CGN=\CGNhat$, then so are
\begin{alignat}{5}
\thetaNS &= \hat{\theta}_{\rm GN}, \qquad &\thetaGN &= \hat{\theta}_{\rm NS}
\qquad &\hbox{for} \qquad \CNS &=  \CGNhat, \qquad &\CGN &=  \CNShat, \label{symmetrySol1} \\
\thetaNS &= \frac{\pi}{2}-\thetaNShat, \qquad &\thetaGN &= \frac{\pi}{2}-\thetaGNhat
\qquad &\hbox{for} \qquad \CNS &= -\CNShat, \qquad &\CGN &= -\CGNhat, \label{symmetrySol2} \\
\thetaNS &= \frac{\pi}{2}-\thetaGNhat, \qquad &\thetaGN &= \frac{\pi}{2}-\thetaNShat
\qquad &\hbox{for} \qquad \CNS &= -\CGNhat, \qquad &\CGN &= -\CNShat. \label{symmetrySol3}
\end{alignat}
In \cref{symmetrySol1}, exchanging $\CNS$ and $\CGN$ leads to an exchange of the $\thetaNS$ and $\thetaGN$ solutions.
In \cref{symmetrySol2}, changing $\CNS$ and $\CGN$ from planar to homeotropic alignment, or vice versa, leads to a rotation of $\thetaNS$ and $\thetaGN$ by $\pi/2$.
In \cref{symmetrySol3}, exchanging $\CNS$ and $\CGN$ (as in \cref{symmetrySol1}) and changing them from planar to homeotropic alignment, or vice versa (as in \cref{symmetrySol2}), leads to an exchange of the $\thetaNS$ and $\thetaGN$ solutions and a rotation of $\pi/2$.

\subsection{Discontinuous solutions for $\theta$}
\label{sec:discontinuous}

Since there are no $x$-derivatives of $\theta$ in the system \cref{thinNS,thinGN,thinYL,thinBC2,thinArea2}, the solution for the director angle $\theta$ can have discontinuities at any number of positions $x$, provided that \cref{thinNS,thinGN,thinYL} are satisfied through each discontinuity and, in particular, since we anticipate on physical grounds that the solutions for the height of the ridge $h$ will be continuous, provided that $\thetaGN-\thetaNS$ is continuous.
Specifically, at any position $x$ there may be discontinuous jumps between a solution $h=\hat{h}$, $p_0=\hat{p}_0$, and $\theta=\hat{\theta}$ and another solution $h=\hat{h}$, $p_0=\hat{p}_0$, and $\theta=q\pi\pm\hat{\theta}$ for any integer $q$.
Discontinuities in $\theta$ correspond to disclination lines, at which, as described in \cref{sec:contactlines},
a description of the nematic only in terms of the director is no longer valid, and
large elastic distortions give rise to a local increase in the elastic energy density and a reduction in the orientational order about the director \cite{SchopohlSluckin1987}.
Although the system \cref{thinNS,thinGN,thinYL,thinBC2,thinArea2} allows for the occurrence of discontinuities in $\theta$, we assume that solutions with disclination lines always have higher energy than those without disclination lines and are therefore always energetically unfavourable \cite{TsakonasDavidsonBrownMottram2007,LewisGarleaAlvaradoetaal2014,WaltonMcKayMottram2018}.
The exception is for situations in which the symmetry of the system necessitates the occurrence of a disclination line at the centreline of the ridge (\ie at $x=0)$.
Indeed, previous work on
nematic droplets by Kleman \cite{KlemanBook1983},
nematic flow in confined microfluidic channels by Sungupta \etal \cite{Sengupta2011},
nematics in a sinusoidal groove by Ohzono and Fukuda \cite{Ohzono2012}, and
nematic mixtures in  ridges by Bao \etal \cite{Bao2021}
show that disclination lines are often found experimentally at or near to the centre of grooves, channels, ridges, and droplets, respectively.
Furthermore, the physical experiments for a pinned static ridge of 5CB described in \cref{sec:experiments} also reveal the presence of a disclination line near to the centreline of the ridge.
In the present work we therefore allow for the occurrence of a discontinuity in $\theta$ at the centreline of the ridge.

Numerical investigations of the system \cref{thinNS,thinGN,thinYL,thinBC2,thinArea2} (see \cref{app:num} for details) suggest that
(i) all solutions for the height of the ridge $h$ decrease monotonically from a maximum value at $x=0$, and have even symmetry about $x=0$,
(ii) all continuous solutions for the director angle $\theta$ also have even symmetry about $x=0$, and
(iii) solutions for $\theta$ with a discontinuity only at $x=0$ have odd symmetry about $x=0$.

In the light of these numerical results, we henceforth restrict our attention to continuous solutions for $\theta$ and $h$ in the right-hand half of the ridge (\ie in $0\le x\le 1$).
Thus the contact-line conditions \cref{thinBC2} are replaced by a single contact-line condition at the right-hand contact line, namely
\begin{align}\label{thinBC}
h = 0 \quad \text{at} \quad x = 1,
\end{align}
and a symmetry and regularity condition at the centreline of the ridge, namely
\begin{align}\label{thinsym}
h_x = 0 \quad \text{at} \quad x = 0.
\end{align}
Also, the area constraint \cref{thinArea2} may be expressed as
\begin{align}\label{thinArea}
1 = 2\int_0^1 h \, \dd x,
\end{align}
and, for simplicity, the right-hand contact angle is hereafter denoted by $\beta=\betaR$.
Once obtained, these solutions in the right-hand half of the ridge can be used to construct solutions for the entire ridge.

\subsection{The limits of strong and weak elasticity}
\label{sec:limits}

Although, as mentioned in \cref{sec:thin}, in what follows we focus on the most interesting regime in which surface-tension, anchoring, and bulk elasticity effects are all comparable, it is useful to discuss briefly the behaviour in the limits in which bulk elastic effects are either much stronger than or much weaker than anchoring effects.
The appropriate parameters for comparing these effects are ratios of the bulk elasticity and the anchoring strength of each interface in the form $K/{\vert\CNS\vert}^{3/2}$ and $K/{\vert\CGN\vert}^{3/2}$.
When elastic effects are much stronger than anchoring effects, \ie in the limit $K/{\vert\CNS\vert}^{3/2}\to \infty$ and $K/{\vert\CGN\vert}^{3/2}\to \infty$, the system \cref{thinNS,thinGN,thinYL,thinBC,thinsym,thinArea} has a solution for the director angle that is uniform everywhere with $\theta\equiv\thetaNS=\thetaGN={\rm constant}$, which will be discussed in \cref{sec:uniform}.
When elastic effects are much weaker than anchoring effects, \ie in the limit $K/{\vert\CNS\vert}^{3/2}\to 0$ and $K/{\vert\CGN\vert}^{3/2}\to 0$, $\thetaNS$ and $\thetaGN$ are constants which are determined by the preferred orientation of the director on the nematic--substrate interface and gas--nematic interface, respectively.
For non-antagonistic anchoring, this again leads to a solution for the director angle that is uniform everywhere with $\theta\equiv\thetaNS=\thetaGN$.
However,
for antagonistic anchoring, this leads to a solution for the director angle that is distorted with $\thetaNS\neq\thetaGN$, except close to the contact line (\ie near to $x=1$), where it can be shown that there are narrow reorientational boundary layers in $\thetaNS$ and $\thetaGN$ of width $\Oh(K/{\vert\CNS\vert}^{3/2})$ and $\Oh(K/{\vert\CGN\vert}^{3/2})$, respectively, in which elastic effects become significant and $\thetaNS$ and $\thetaGN$ approach the same constant value determined by the interface with the stronger anchoring.

\section{Uniform director solutions}
\label{sec:uniform}

We now consider uniform director solutions to the system \cref{thinNS,thinGN,thinYL,thinBC,thinsym,thinArea}.
This system has uniform director solutions valid for all values of $K$, $\CNS$, and $\CGN$ given by
\begin{align}\label{uniform}
\thetaNS=\thetaGN\equiv\dfrac{m\pi}{2}, \quad h = \hiso(x), \quad \text{and} \quad p_0 = \pI,
\end{align}
so that $\theta\equiv m\pi/2$, where $m=0$ or $m=1$, and $\hiso$ and $\pI$ are the solutions for the height and the Lagrange multiplier for an isotropic ridge given by \cref{hiso}, respectively.
In particular, the maximum height at the middle of the ridge and the contact angle are given by $\hmax=3/4$ and $\beta=3/2$, respectively.
We note that these solutions are independent of $K$, $\CNS$, and $\CGN$; however, the second-order energy given by \cref{thinEtot,thinEparts} of these solutions does depend on $\CNS$ and $\CGN$ and is given by
\begin{equation}\label{energyU}
\DelE = \dfrac{3}{2}+\left(\CNS+\CGN\right)\cos m\pi.
\end{equation}
Inspection of \cref{energyU} shows that
a uniform solution aligned everywhere with the homeotropic anchoring,
which we term \emph{an $\Hsol$ solution},
with $\theta\equiv\thetaNS=\thetaGN \equiv \pi/2$ (\ie $m=1$) and hence $\n\equiv\unitz$,
is energetically preferred when $\CNS+\CGN>0$, and
a uniform solution aligned everywhere with the planar anchoring,
which we term \emph{a $\Psol$ solution},
with $\theta\equiv\thetaNS=\thetaGN \equiv 0$ (\ie $m=0$) and hence $\n\equiv\unitx$,
is energetically preferred when $\CNS+\CGN<0$.
When $\CNS+\CGN>0$ either the anchoring is non-antagonistic with homeotropic anchoring on both interfaces or
the anchoring is antagonistic with the homeotropic anchoring of one interface being stronger than the planar anchoring of the other interface.
Similarly,
when $\CNS+\CGN<0$ either the anchoring is non-antagonistic with planar anchoring on both interfaces or
the anchoring is antagonistic with the planar anchoring of one interface being stronger than the homeotropic anchoring of the other interface.
For future reference, we note that from \cref{energyU} the $\Hsol$ and $\Psol$ solutions have second-order energies $\DelEH$ and $\DelEP$ given by
\begin{align}\label{EHP}
\DelEH=\dfrac{3}{2}-\CNS-\CGN \quad \text{and} \quad \DelEP=\dfrac{3}{2}+\CNS+\CGN,
\end{align}
respectively.

\begin{figure}[tp]
\begin{center}
\begin{tabular}{cc}
\includegraphics[width=0.45\linewidth]{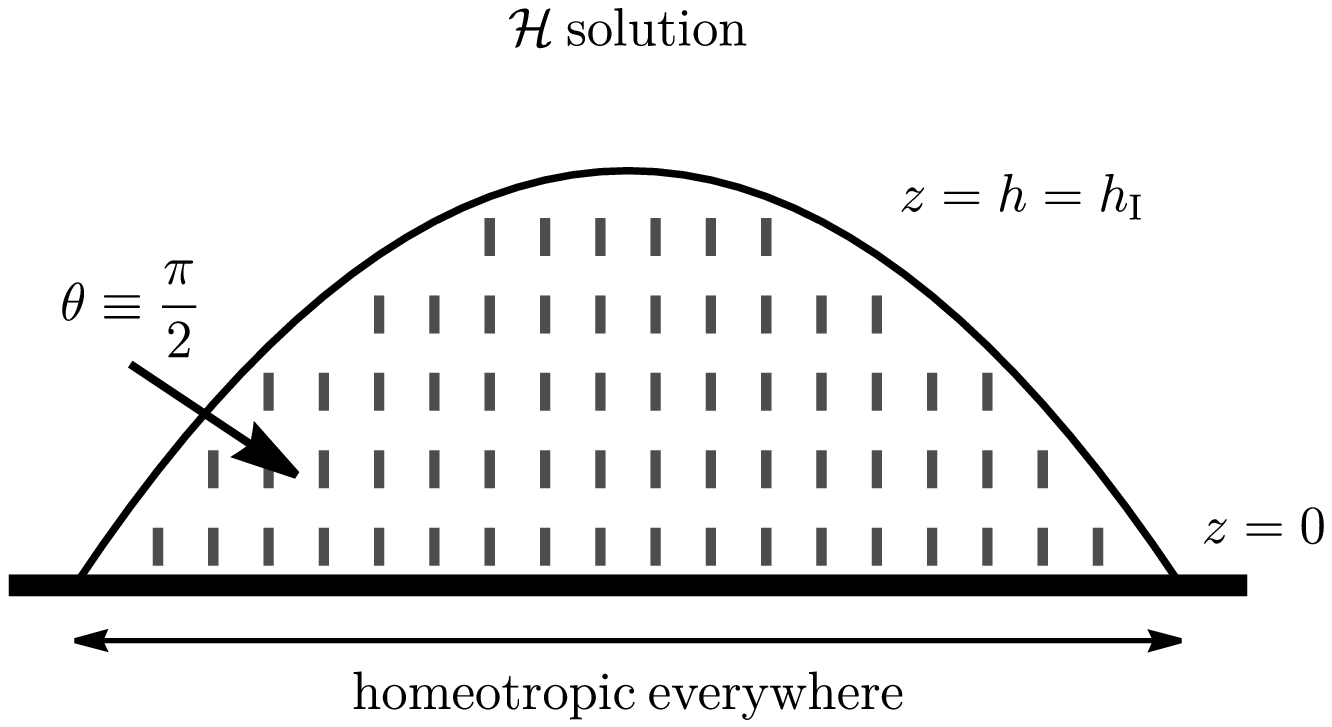} &
\includegraphics[width=0.45\linewidth]{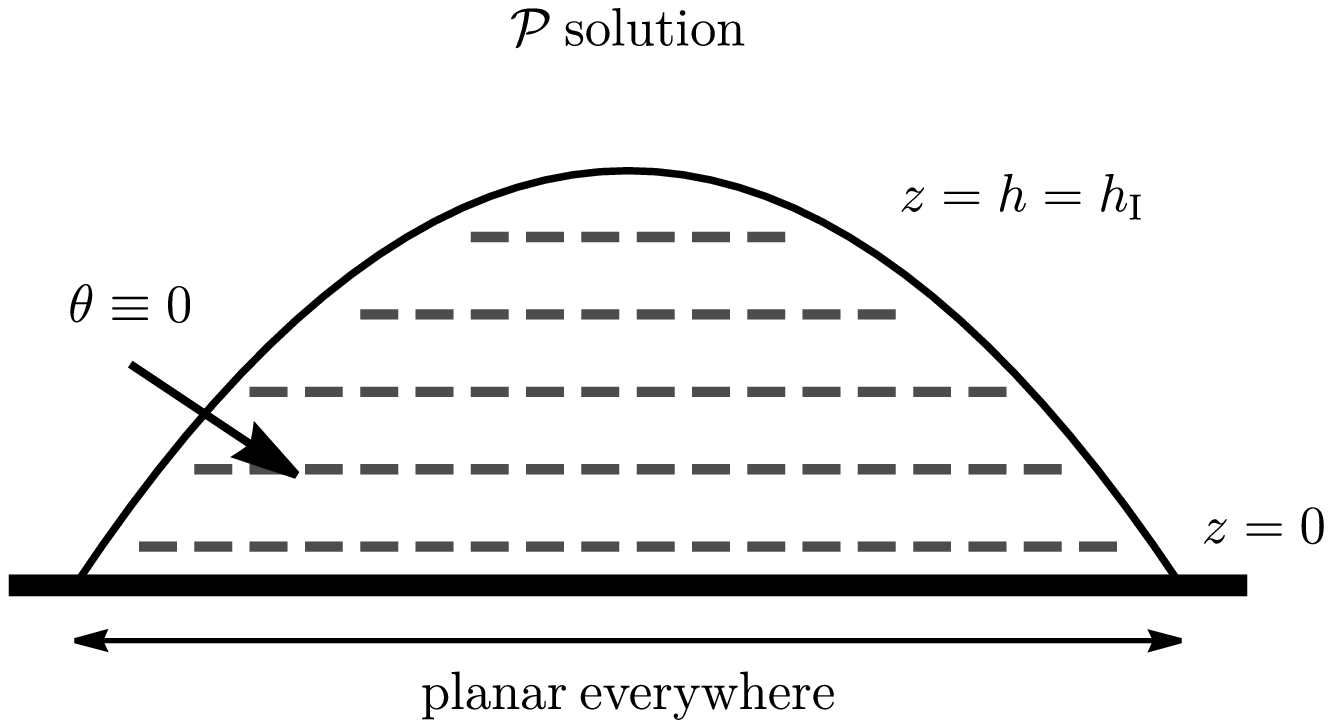} \\
(a) & (b) \\
\end{tabular}
\end{center}
\caption{
Sketches of (a) an $\Hsol$ solution and (b) a $\Psol$ solution.
The director field $\n$ is indicated by the dark grey rods, with $\n\equiv\unitz$ in (a) and $\n\equiv\unitx$ in (b).
}
\label{fig:2}
\end{figure}

\cref{fig:2} shows sketches of
an $\Hsol$ solution (in which the director is aligned everywhere with the homeotropic anchoring) and
a $\Psol$ solution (in which the director is aligned everywhere with the planar anchoring).

Finally,
we note that although the uniform director solutions are always solutions of the system \cref{thinNS,thinGN,thinYL,thinBC,thinsym,thinArea},
as we shall see later,
they are not always the energetically preferred solutions of the system.

\section{Distorted director solutions}
\label{sec:distorted}

We now consider distorted director solutions to the system \cref{thinNS,thinGN,thinYL,thinBC,thinsym,thinArea}.
In addition to the uniform director solutions with $\thetaNS=\thetaGN$ valid for all values of $K$, $\CNS$ and $\CGN$ described in \cref{sec:uniform},
for certain values of $K$, $\CNS$, and $\CGN$ this system also has distorted director solutions.
For these distorted director solutions, the anchoring conditions \cref{thinNS} and \cref{thinGN} can be combined and rearranged into the form
\begin{align}\label{t1t2neq}
\dfrac{\hc}{h} &= \dfrac{\sin 2\thetaNS-\sin 2\thetaGN}{2\left(\thetaNS-\thetaGN\right)}
\quad \hbox{for} \quad \thetaNS\ne\thetaGN,
\end{align}
where
\begin{align}\label{hc}
\hc &= K\left(\dfrac{1}{\CNS}+\dfrac{1}{\CGN}\right)
\end{align}
is the Jenkins--Barratt--Barbero--Barberi critical thickness \cite{JenkinsBarratt1974,BarberoandBarberi1983} discussed in \cref{sec:hc} non-dimensionalised with $\epsilon \td$.
We note that although $\hc$ is referred to as a thickness it can be positive, negative or zero.
\cref{tab:1} summarises how the sign of $\hc$ depends on the values of $\CNS$ and $\CGN$, together with the corresponding type of anchoring and the nature of the stronger anchoring.

\begin{table}[tp]
\centering
\begin{tabular}{|c|c|c|c|}
\hline
$\CNS$ and $\CGN$ & \quad Sign of $\hc$ \quad & \quad Type of Anchoring \quad & \quad Stronger Anchoring \quad \\
\hline
\hline
$\CNS\ge\CGN>0$ & \multirow{2}{*}{$\hc>0$} & \multirow{2}{*}{non-antagonistic} & \multirow{2}{*}{homeotropic} \\
\cline{1-1}
$\CGN>\CNS>0$ & & & \\
\hline
$\CNS\le\CGN<0$ & \multirow{2}{*}{$\hc<0$} & \multirow{2}{*}{non-antagonistic} & \multirow{2}{*}{planar} \\
\cline{1-1}
$\CGN<\CNS<0$ & & & \\
\hline
$\CNS=-\CGN$ & $\hc=0$ & antagonistic & equal \\
\hline
$\CNS>-\CGN>0$ & \multirow{2}{*}{$\hc<0$} & \multirow{2}{*}{antagonistic} & \multirow{2}{*}{homeotropic} \\
\cline{1-1}
$\CGN>-\CNS>0$ & & & \\
\hline
$\CNS<-\CGN<0$ &  \multirow{2}{*}{$\hc>0$} & \multirow{2}{*}{antagonistic} & \multirow{2}{*}{planar} \\
\cline{1-1}
$\CGN<-\CNS<0$ & &  & \\
\hline
\end{tabular}
\caption{
A summary of how the sign of $\hc$ depends on the values of $\CNS$ and $\CGN$, together with the corresponding type of anchoring and the nature of the stronger anchoring.
}
\label{tab:1}
\end{table}

Equations \cref{t1t2neq} and \cref{hc} give information about the range of values of $\hc$, and therefore the ranges of values of $K$, $\CNS$, and $\CGN$, for which distorted director solutions are possible.
In particular, using the Mean Value Theorem shows that the magnitude of the right-hand side of \cref{t1t2neq} is less than unity, and therefore that $h>\vert\hc\vert$ when $\thetaNS\neq\thetaGN$.
Thus, the director can be distorted in the $z$-direction, \ie can have a solution with $\thetaNS\neq\thetaGN$, only at positions $x$ for which $h>\vert\hc\vert$.
On the other hand, at positions $x$ for which $h\le\vert\hc\vert$ the director can only be uniform and hence is given by \cref{uniform}.
Consequently, when $h\le\vert\hc\vert$ for all $0 \le x \le 1$, the solution is either an $\Hsol$ or a $\Psol$ solution.
On the other hand, when $h>\vert\hc\vert$ at $x=0$, either the director is uniform with $\thetaNS=\thetaGN$ in a \emph{uniform region} $\xc \le x \le 1$ in which $h\le\vert\hc\vert$, where $x=\xc$ is the position at which $h(\xc)=\vert\hc\vert$, and distorted with $\thetaNS\neq\thetaGN$ in a \emph{distorted region} $0 \le x < \xc$ in which $h>\vert\hc\vert$, or it is uniform everywhere, \ie an $\Hsol$ or a $\Psol$ solution given by \cref{uniform}.
We term
a distorted director solution with
a uniform region in which the director is aligned with the homeotropic anchoring
\emph{a $\DHsol$ solution}
and
a distorted director solution with
a uniform region in which the director is aligned with the planar anchoring
\emph{a $\DPsol$ solution}.
In summary, $\DHsol$ and $\DPsol$ solutions are given by
\begin{align}\label{theta_distort}
\theta =
\begin{cases}
\thetaNS+\left(\thetaGN-\thetaNS\right)\dfrac{z}{h} &\text{for} \quad 0 \le x < \xc,\\
\dfrac{m\pi}{2} &\text{for} \quad \xc \le x \le 1,
\end{cases}
\end{align}
with $m=1$ for a $\DHsol$ solution and $m=0$ for a $\DPsol$ solution.

In the uniform region, the nematic Young--Laplace equation \cref{thinYL} reduces to $h_{xx}=-p_0$, which may be integrated with respect to $x$ subject to the contact-line condition \cref{thinBC} and $h(\xc)=\vert\hc\vert$ to yield the height of the ridge $h$ in terms of the two as-yet-unknown quantities $p_0$ and $\xc$, namely
\begin{align}\label{hxc}
h &= (1-x)\left[\dfrac{p_0}{2}(x-\xc)+\dfrac{\vert\hc\vert}{1-\xc}\right] \quad \text{for} \quad \xc \le x \le 1.
\end{align}
%
In the limit $\xc \to 0^+$, the solution \cref{hxc} must satisfy the symmetry and regularity condition \cref{thinsym} and the area constraint \cref{thinArea}, which yield $\vert\hc\vert=p_0/2$ and $p_0=\pI=3/2$, respectively, and hence $h\to\hiso$ and $\vert\hc\vert \to 3/4$, where $\hiso$ is given by \cref{hiso}.
In particular, in this limit the uniform region occupies the entire ridge, and the $\DHsol$ and $\DPsol$ solutions reduce to the $\Hsol$ and $\Psol$ solutions, respectively.

In the special case in which the anchoring strengths of the nematic--substrate interface and the gas--nematic interface are equal and opposite (\ie when $\CGN=-\CNS$) then $\hc=0$ and $\vert\xc\vert=1$ so that the distorted region occupies the entire ridge and the director is distorted everywhere, \ie $\thetaNS \neq \thetaGN$ for $0 \le x < 1$.
We term such a solution \emph{a $\Dsol$ solution}.
Note that in this special case (and only in this special case) anchoring breaking occurs on \emph{both} interfaces close to the contact line, and the director angle at the contact line is the \emph{mean} of the preferred values on the two interfaces, \ie $\thetaNS=\thetaGN=\pi/4$ at $x=1$.
Further analytical progress can be made for the $\Dsol$ solution in the asymptotic limits $\CGN=-\CNS\to0$ and $\CGN=-\CNS\to\infty$ (see Cousins \cite{Cousinsthesis} for details), but, for brevity, we do not describe this analysis here.

\begin{figure}[tp]
\begin{center}
\begin{tabular}{cc}
\includegraphics[width=0.45\linewidth]{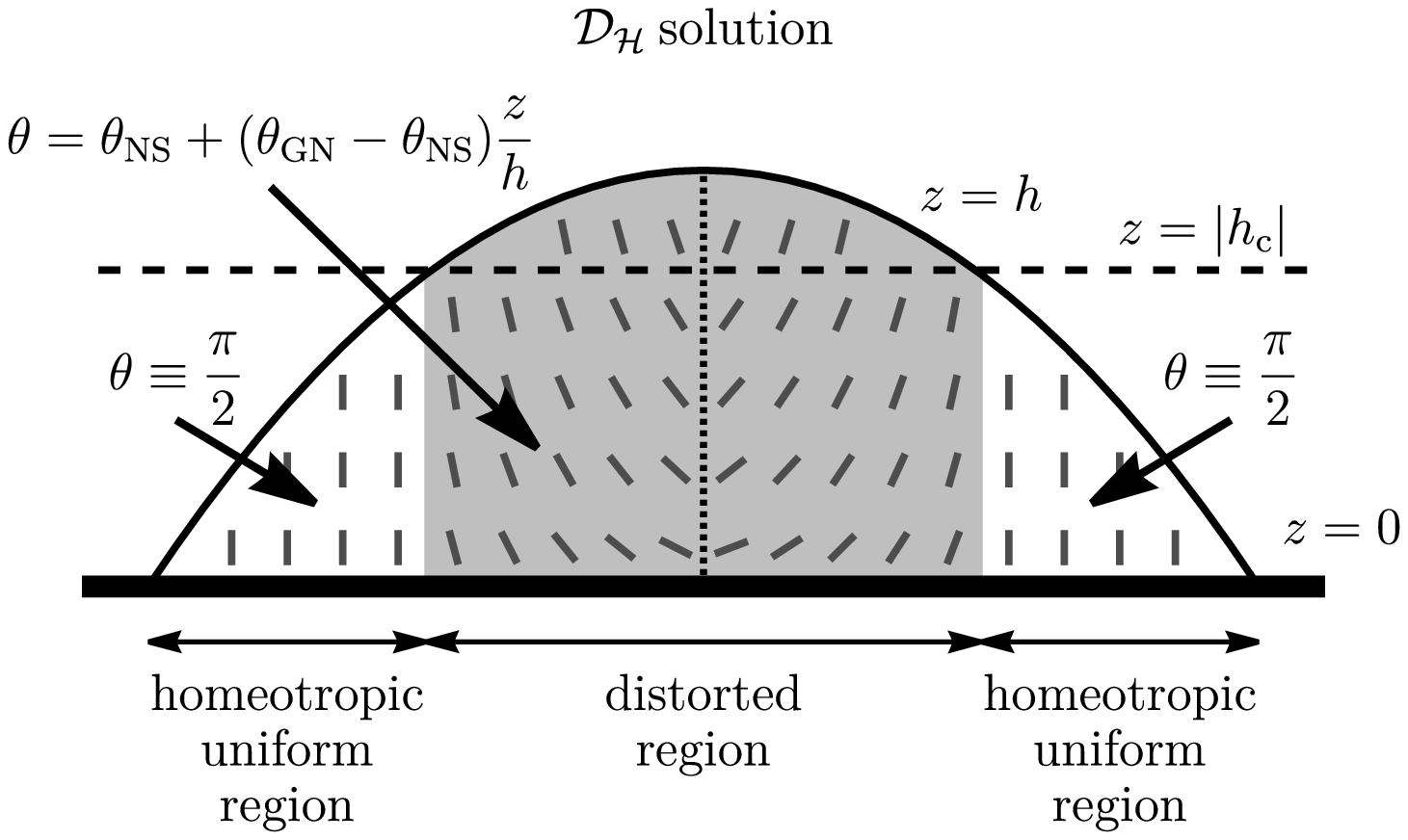} &
\includegraphics[width=0.45\linewidth]{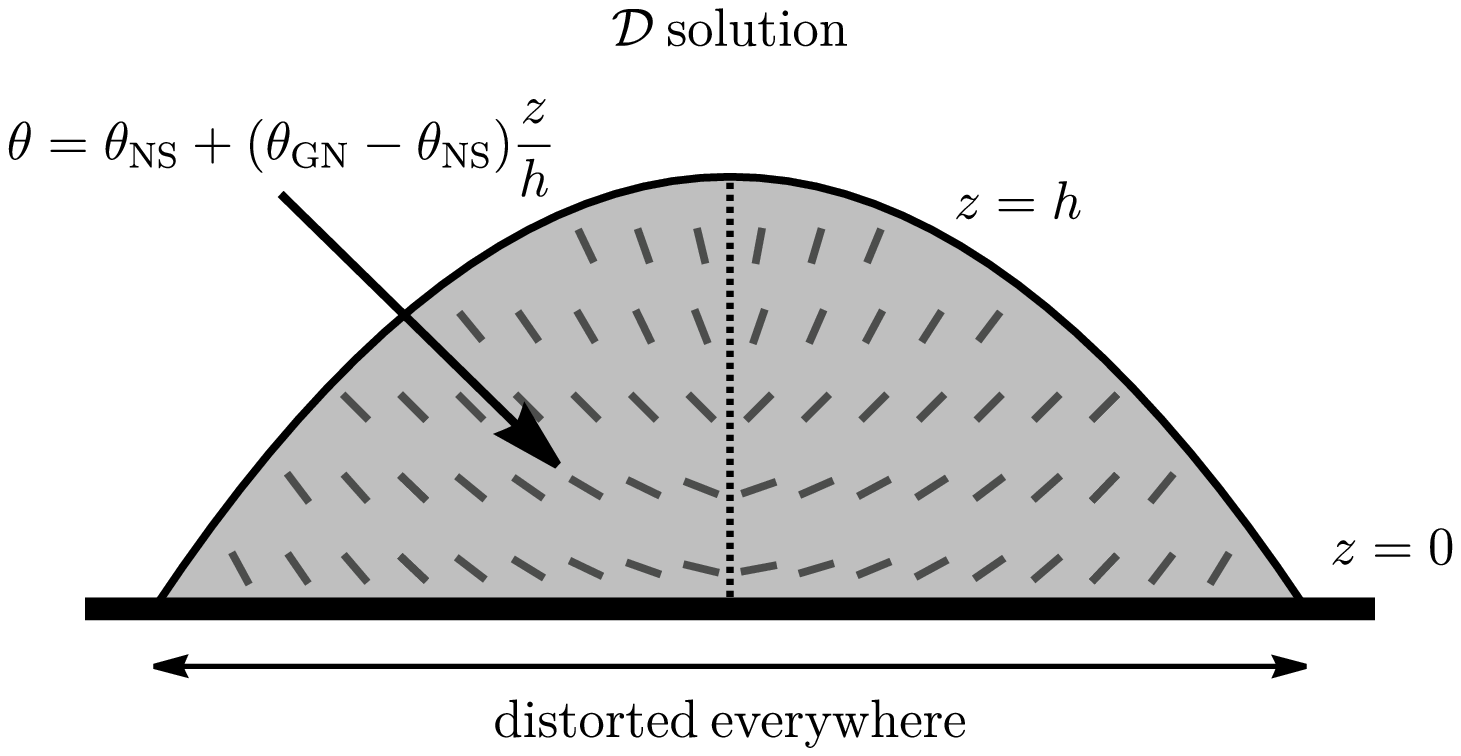}\\[-0.6cm]
(a) & (b) \\[0.1cm]
\multicolumn{2}{c}{\includegraphics[width=0.45\linewidth]{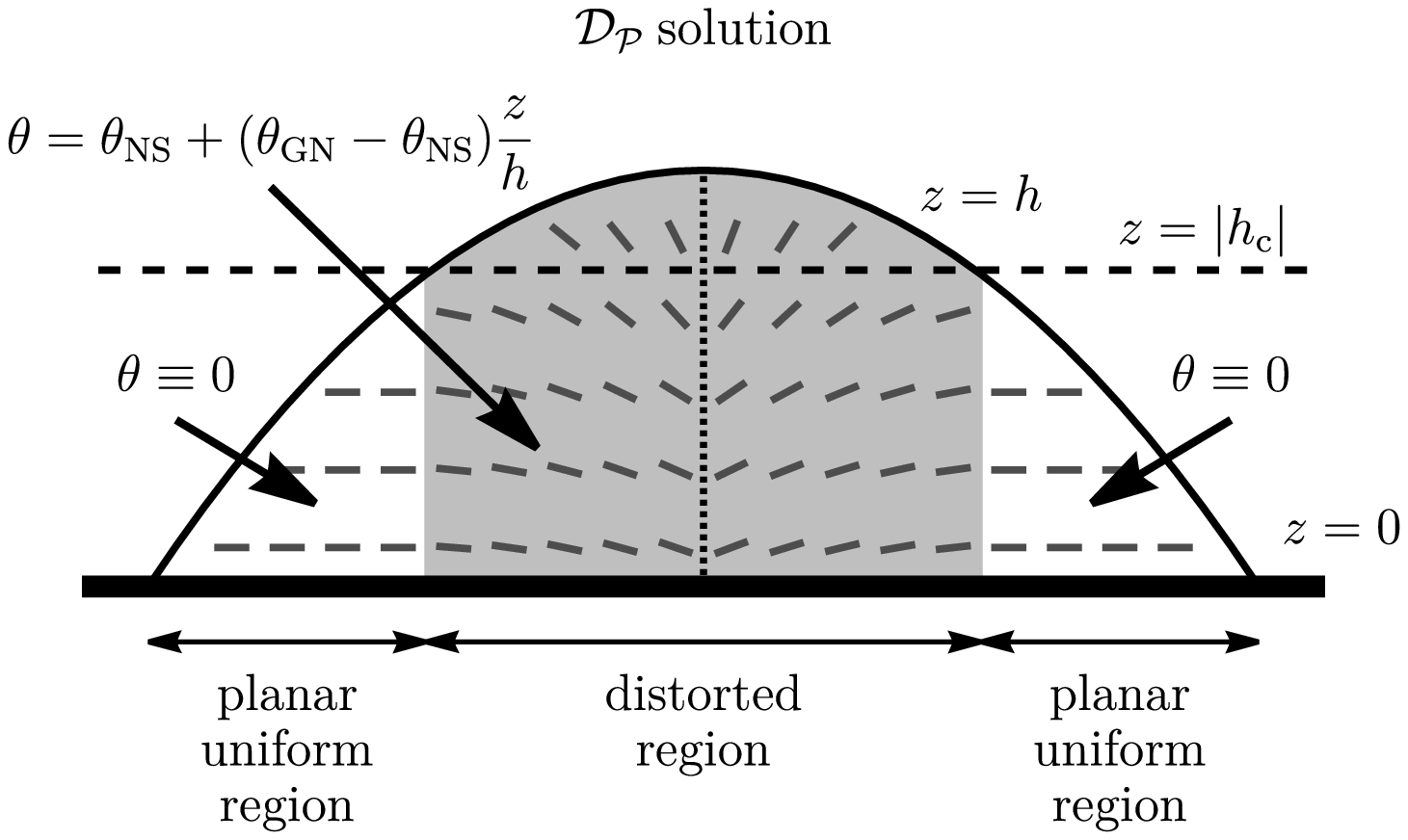}} \\[-0.6cm]
\multicolumn{2}{c}{(c)} \\
\end{tabular}
\end{center}
\caption{
Sketches of (a) a $\DHsol$ solution, (b) a $\Dsol$ solution, and (c) a $\DPsol$ solution.
The director field $\n$ is indicated by the dark grey rods, the critical thickness $z=h(\xc)=\vert\hc\vert$ is shown by a dashed line (note that $z=h(\xc)=\vert\hc\vert=0$ is not visible in (b)), the disclination line at $x=0$ is shown with a dotted line, the uniform regions $\xc \le \vert x\vert \le 1$ in which $h\le\vert\hc\vert$ are indicated in white, and the distorted region $\vert x\vert < \xc$ in which $h>\vert\hc\vert$ is indicated in grey.
}
\label{fig:3}
\end{figure}

\cref{fig:3} shows sketches of a $\DHsol$ solution, a $\Dsol$ solution (in which the director is distorted everywhere), and a $\DPsol$ solution.

\section{Energetically--preferred solutions}
\label{sec:preferred}

We now solve the system \cref{thinNS,thinGN,thinYL,thinBC,thinsym,thinArea} and determine the regions of parameter space in which the five qualitatively different types of solution, namely the $\Hsol$, $\Psol$, $\DHsol$, $\DPsol$, and $\Dsol$ solutions, introduced in \cref{sec:uniform,sec:distorted} and sketched in \cref{fig:2,fig:3}, are energetically preferred.
Specifically, in order to determine which of these solutions has the lowest energy, and which is therefore the energetically-preferred solution, we compare the second-order energy of the distorted director solution, namely $\DelE$ given by \cref{thinEtot,thinEparts}, with the second-order energies of the $\Hsol$ and $\Psol$ solutions, namely $\DelEH$ and $\DelEP$ given by \cref{EHP}.
In general, the director angles $\thetaNS$ and $\thetaGN$ and the height of the ridge $h$ must be determined numerically \cite{Notei}.
The numerical procedure used employed the MATLAB stiff differential-algebraic equation solver \emph{ode15s} \cite{MATLAB2019} (see \cref{app:num} for details).

\subsection{Antagonistic anchoring (\ie $\CNS\CGN<0$)}
\label{sec:antagonistic}

\begin{figure}[p]
\begin{center}
\setlength\tabcolsep{1pt}
\begin{tabular}{ccc}
\includegraphics[width=0.32\linewidth]{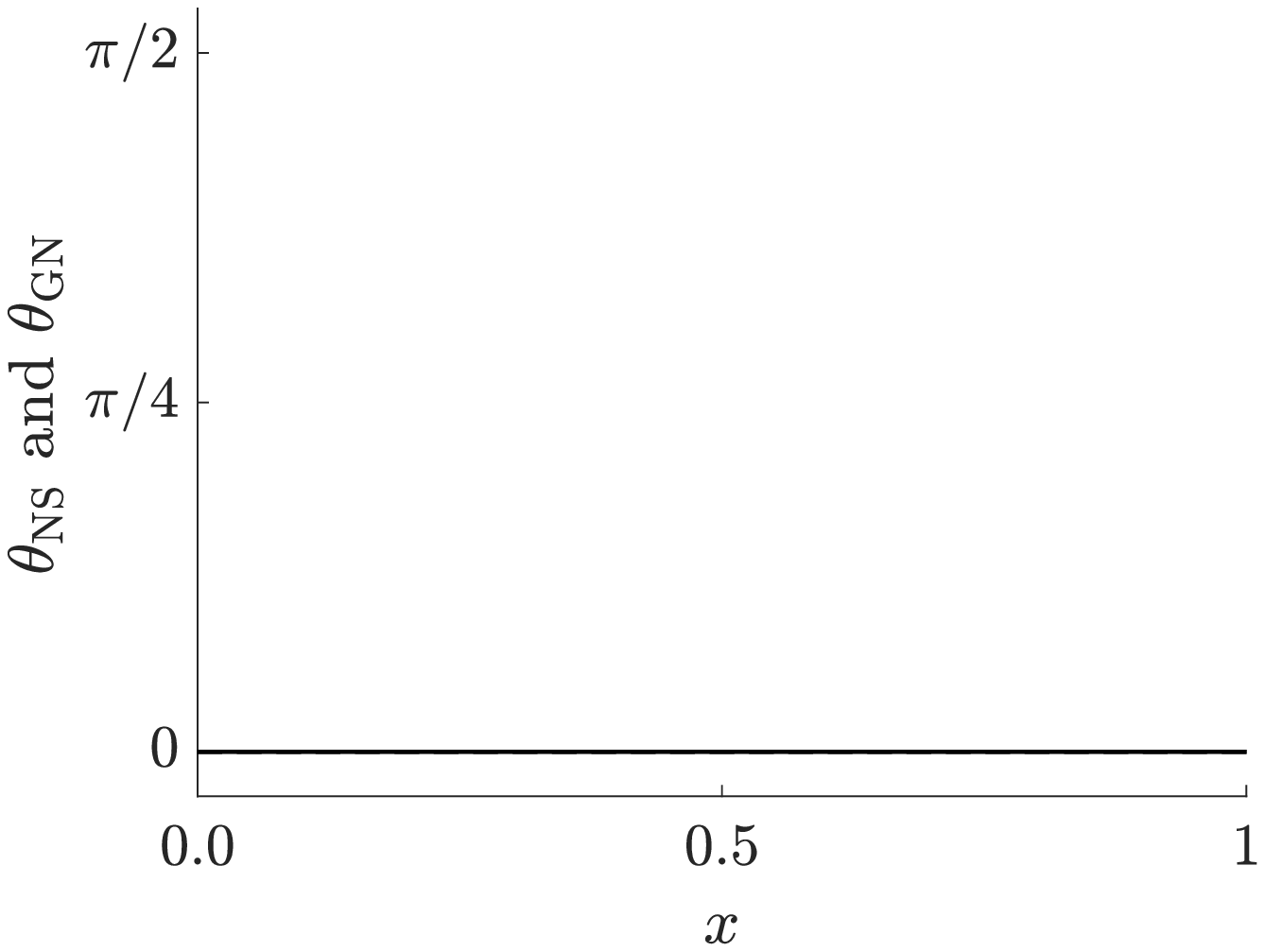} &
\includegraphics[width=0.32\linewidth]{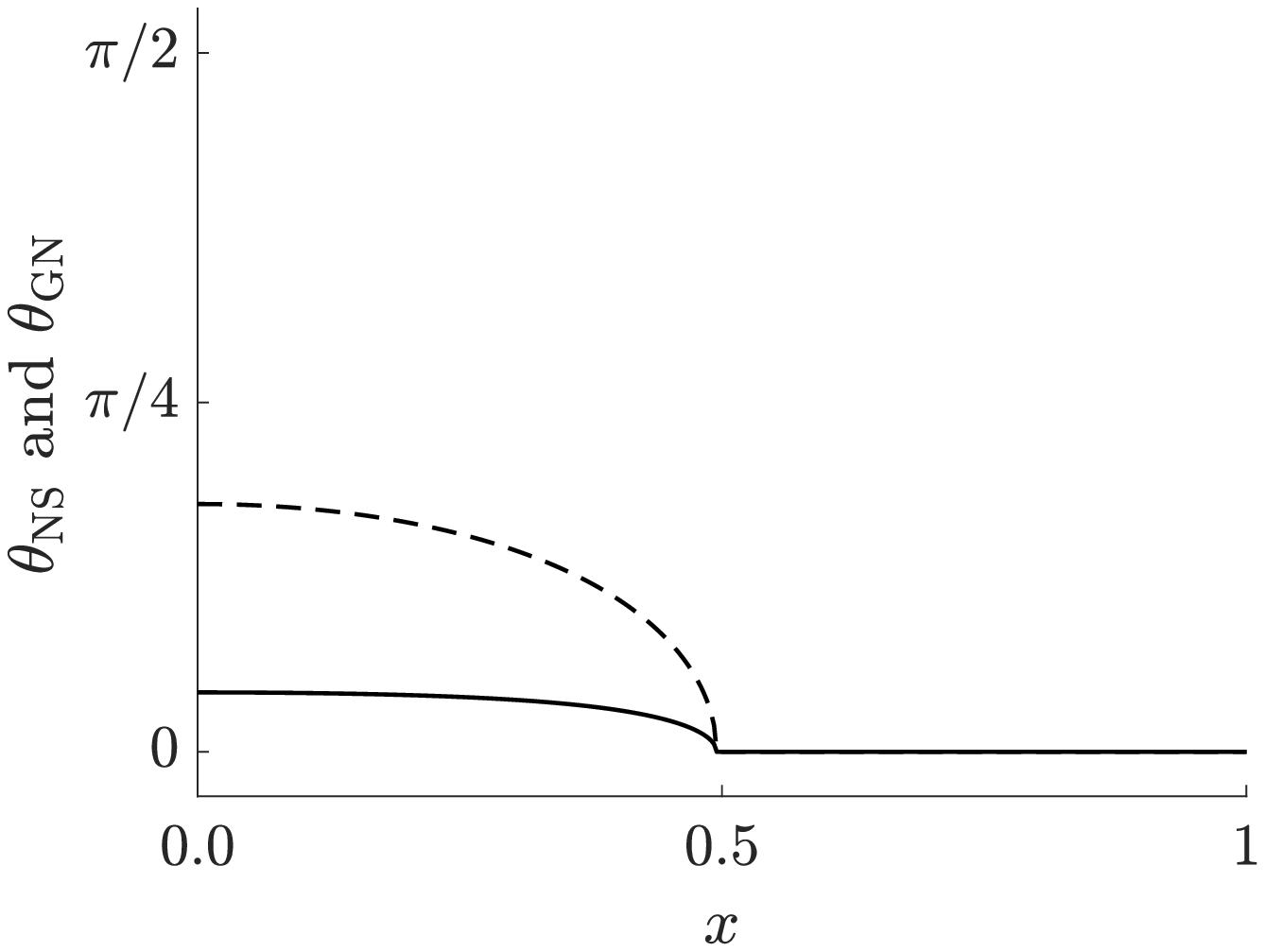} &
\includegraphics[width=0.32\linewidth]{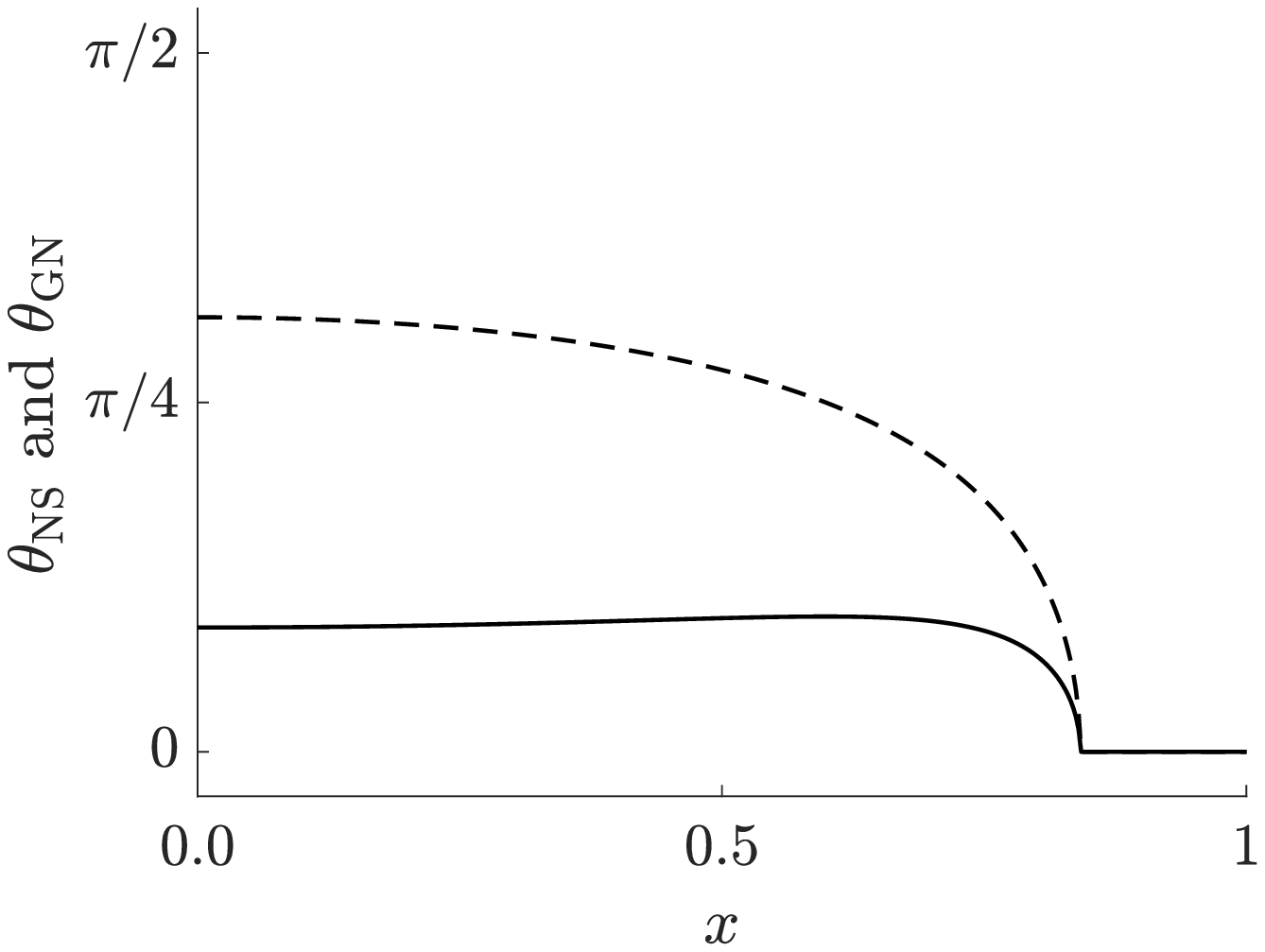} \\
(a) & (b) & (c) \\[1cm]
\includegraphics[width=0.32\linewidth]{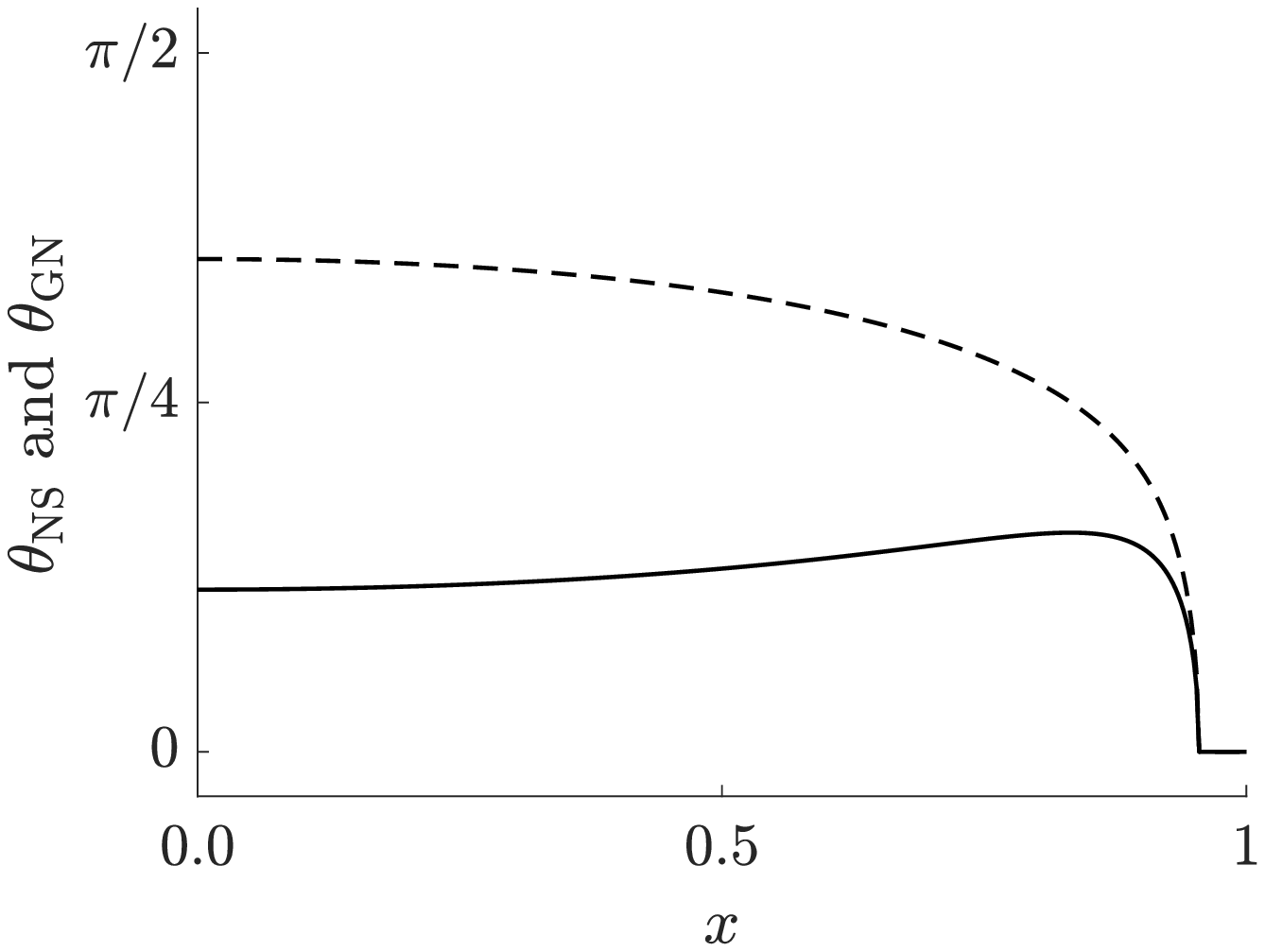} &
\includegraphics[width=0.32\linewidth]{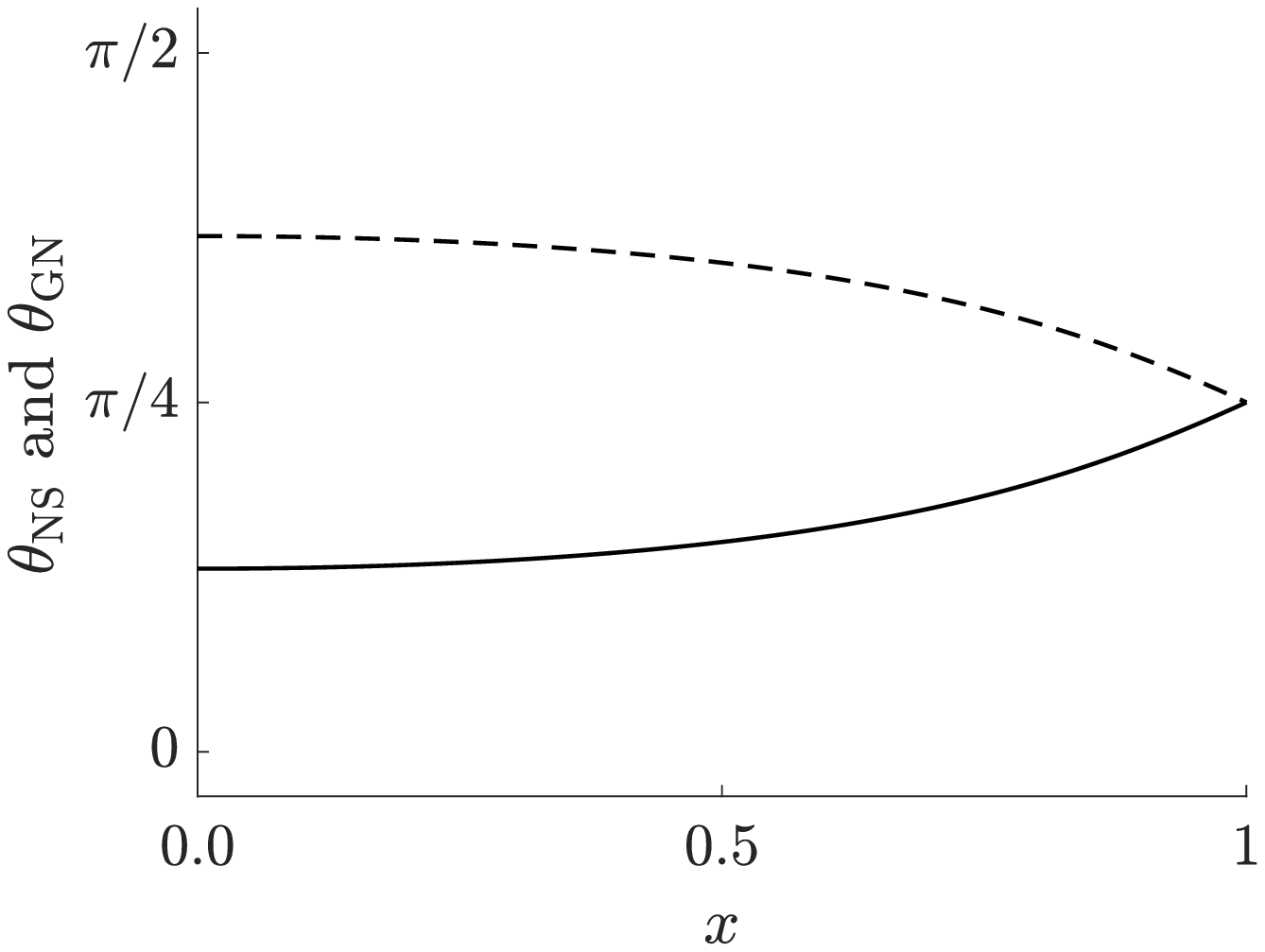} &
\includegraphics[width=0.32\linewidth]{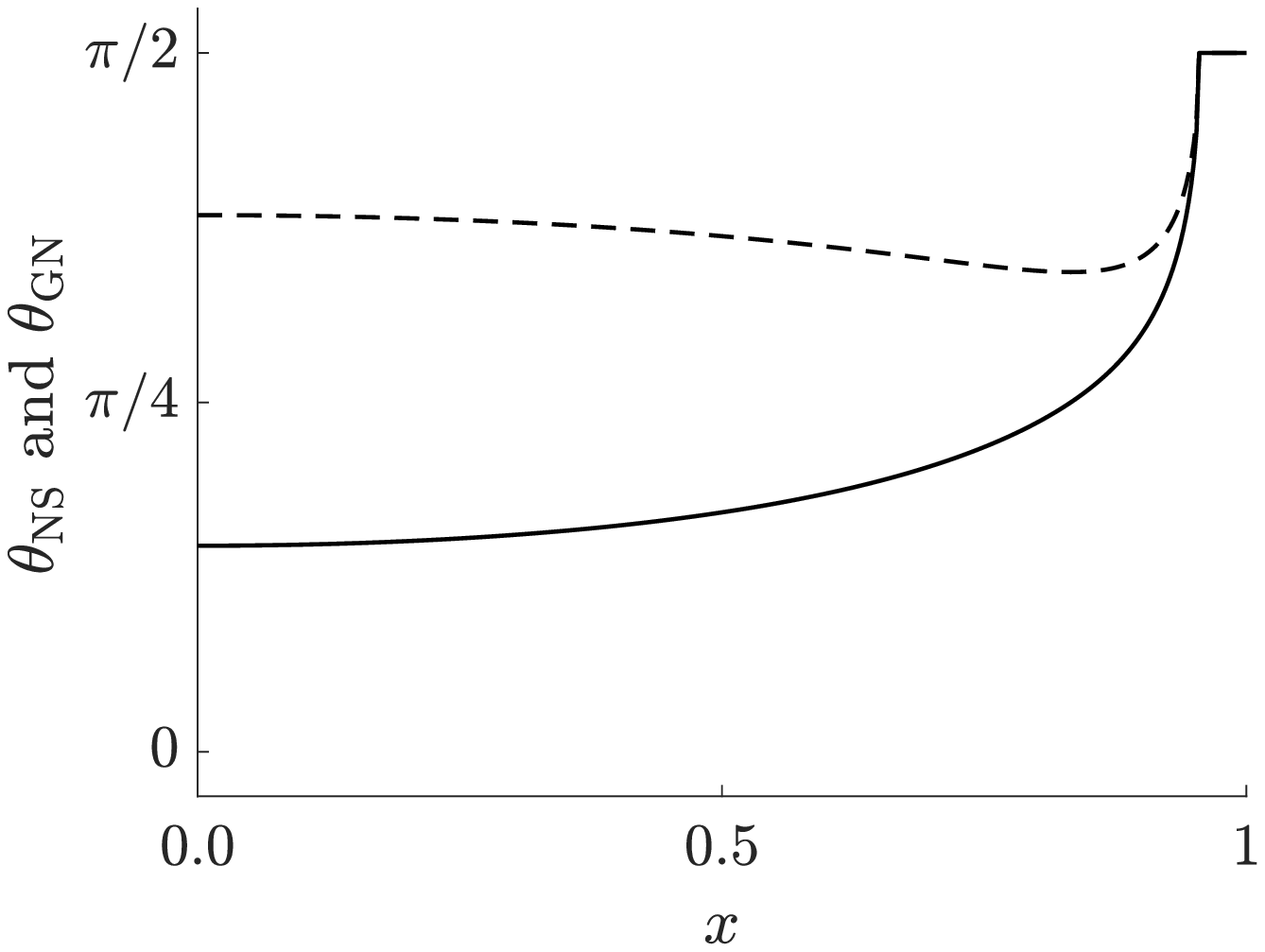} \\
(d) & (e) & (f) \\[1cm]
\includegraphics[width=0.32\linewidth]{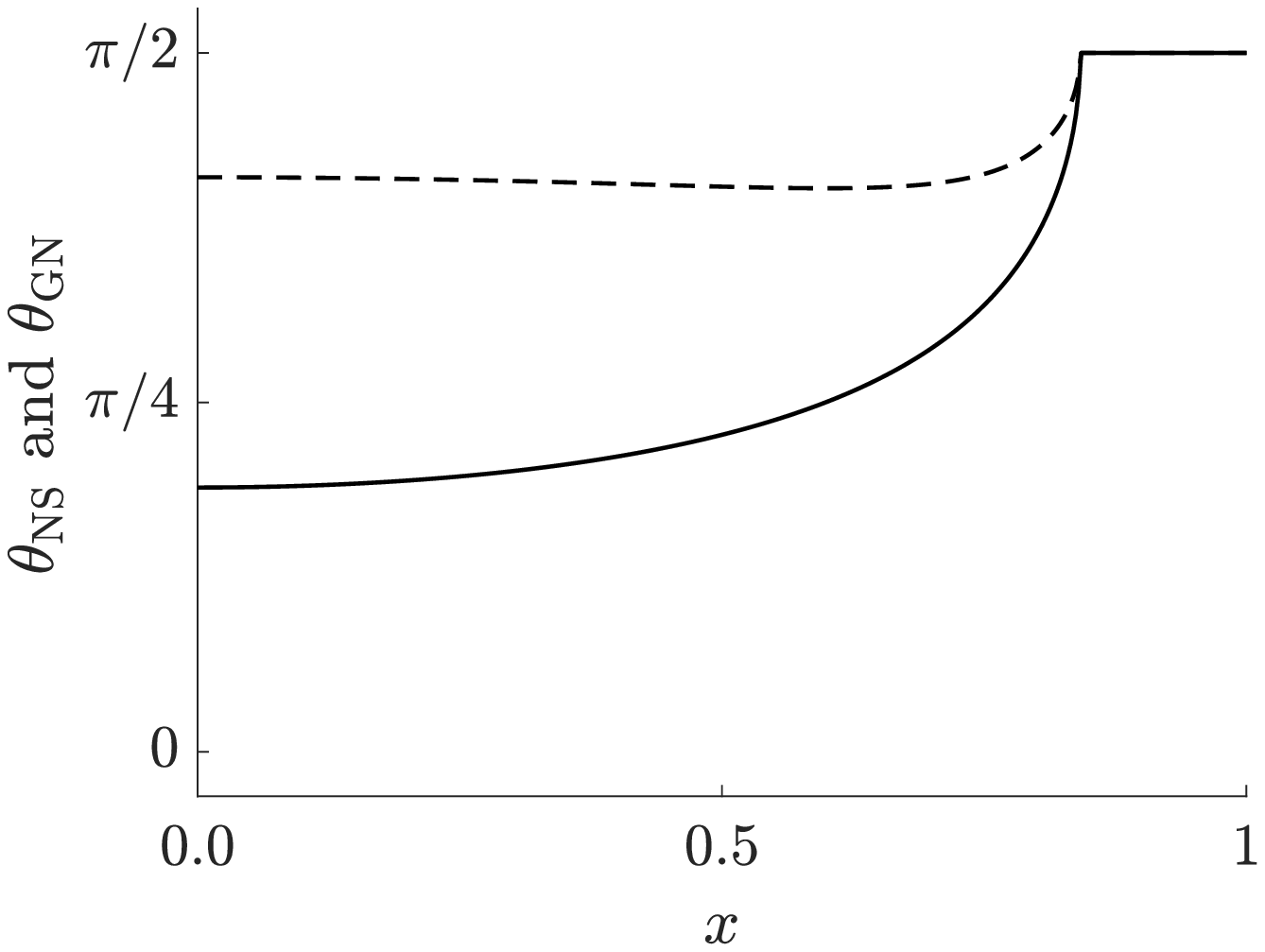} &
\includegraphics[width=0.32\linewidth]{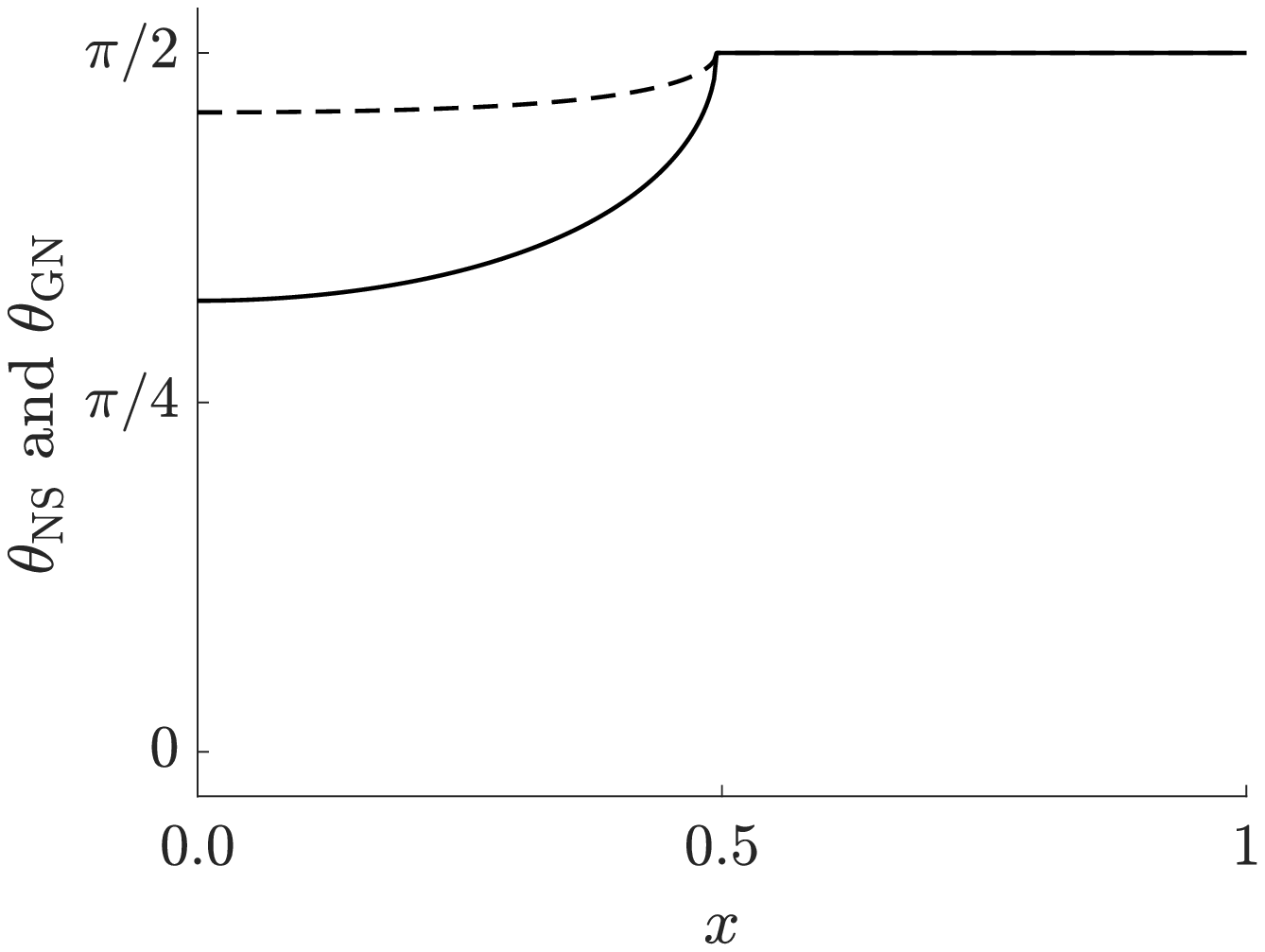} &
\includegraphics[width=0.32\linewidth]{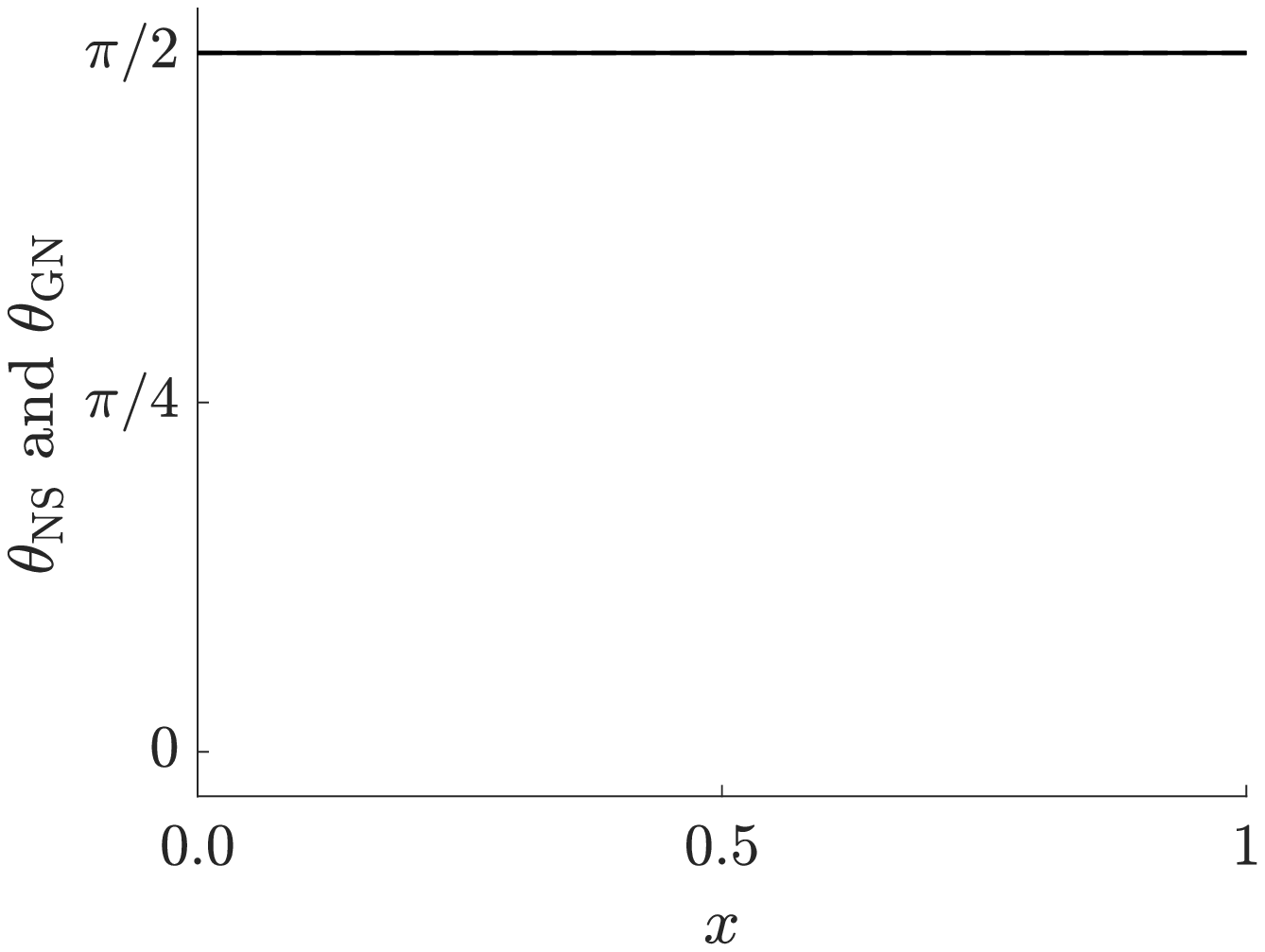} \\
(g) & (h) & (i) \\
\end{tabular}
\end{center}
\caption{
The director angles on the nematic--substrate interface $\thetaNS$ (solid line) and the gas--nematic interface $\thetaGN$ (dashed line) for representative numerically-calculated energetically-preferred solutions when $K=1$ in the case of antagonistic anchoring with homeotropic anchoring on the gas--nematic interface and planar anchoring on the nematic--substrate interface with a constant difference between $\CNS$ and $\CGN=\CNS+5.5>0$ (chosen to show a full range of behaviours in which elasticity and the anchoring of both interfaces are all comparable) for (a) $\CNS=-4.5$, (b) $\CNS=-4.25$, (c) $\CNS=-3.5$, (d) $\CNS=-3$, (e) $\CNS=-2.75$, (f) $\CNS=-2.5$, (g) $\CNS=-2$, (h) $\CNS=-1.25$, and (i) $\CNS=-1$.
}
\label{fig:4}
\end{figure}

\begin{figure}[p]
\begin{center}
\setlength\tabcolsep{1pt}
\begin{tabular}{ccc}
\includegraphics[width=0.32\linewidth]{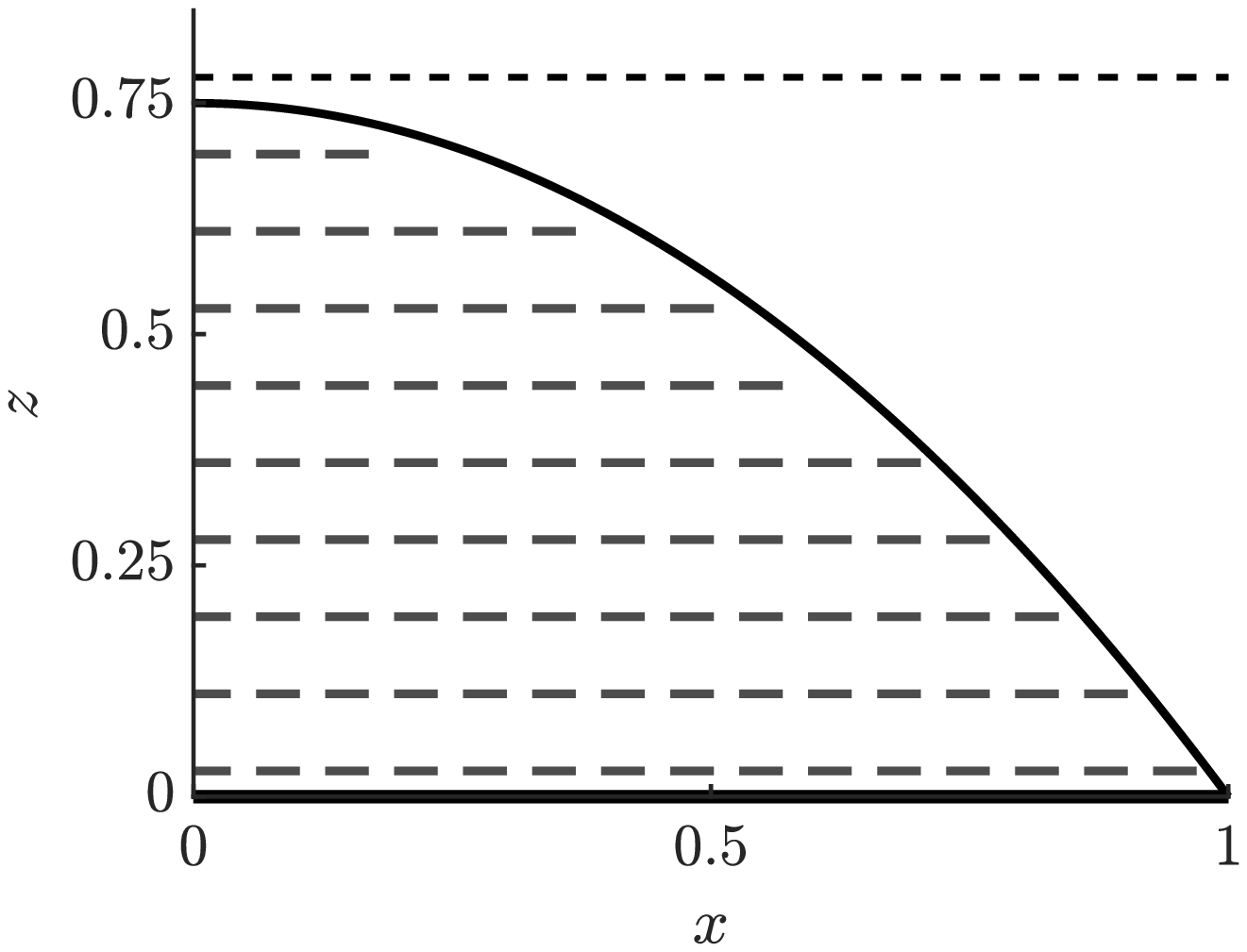} &
\includegraphics[width=0.32\linewidth]{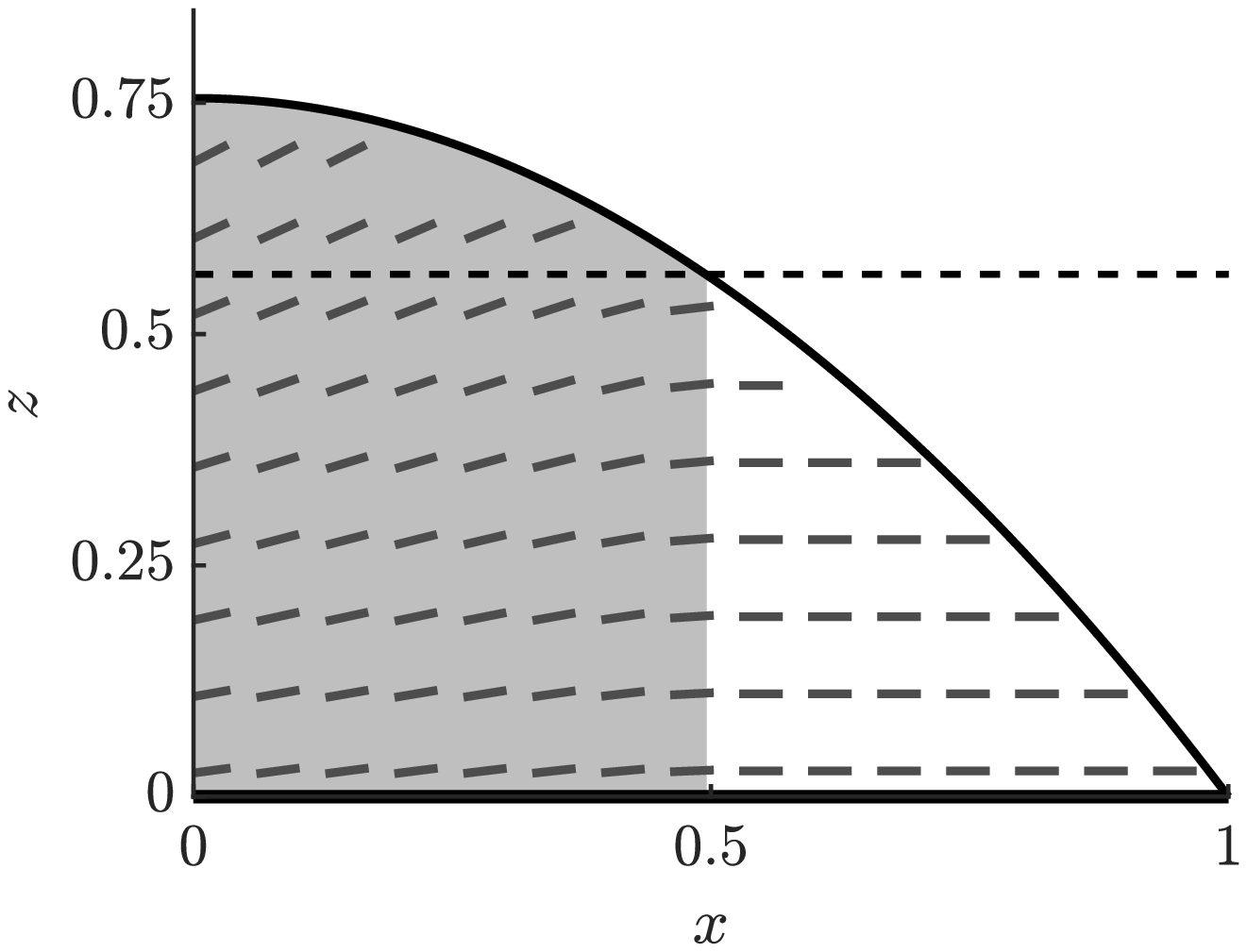} &
\includegraphics[width=0.32\linewidth]{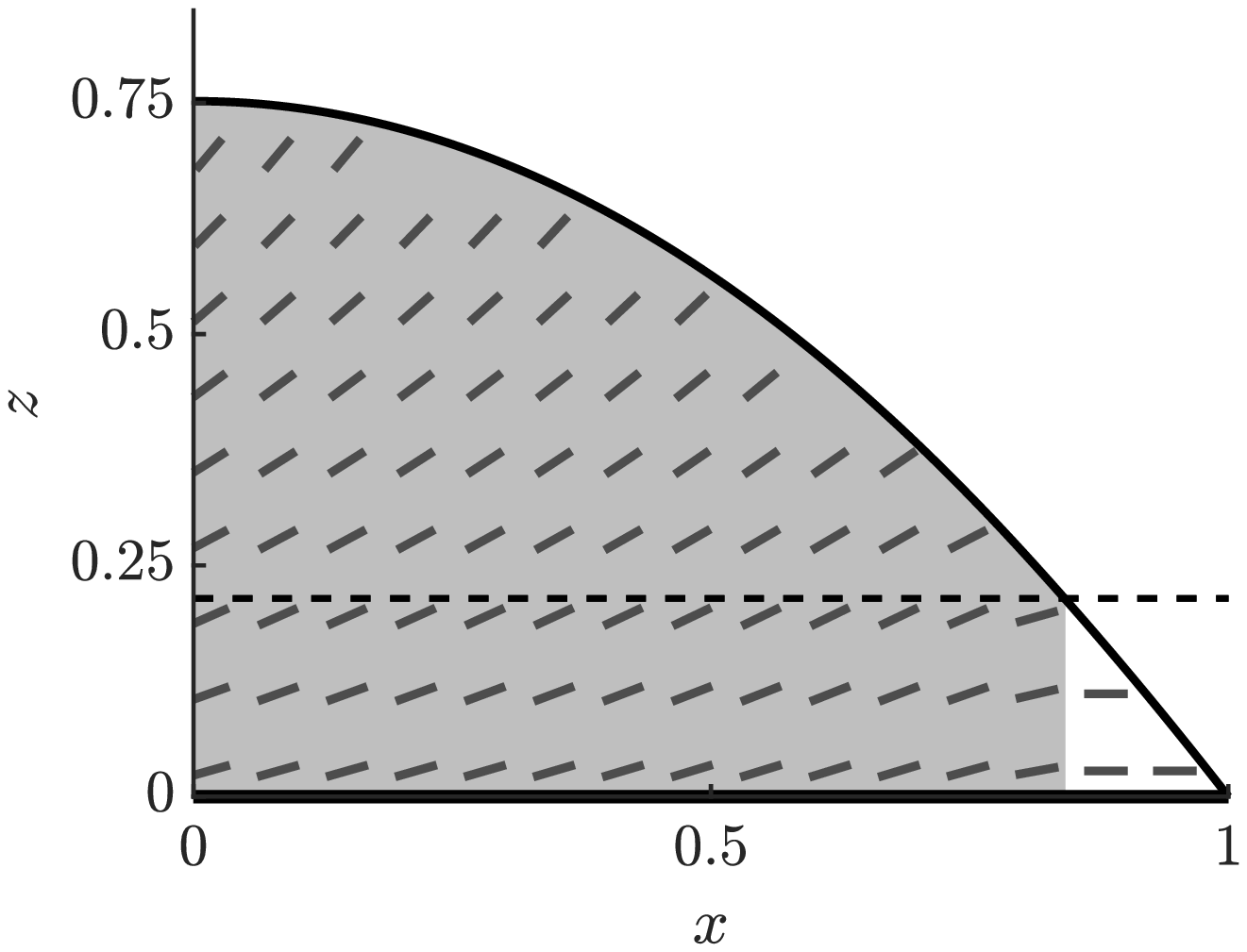} \\
(a) & (b) & (c) \\[1cm]
\includegraphics[width=0.32\linewidth]{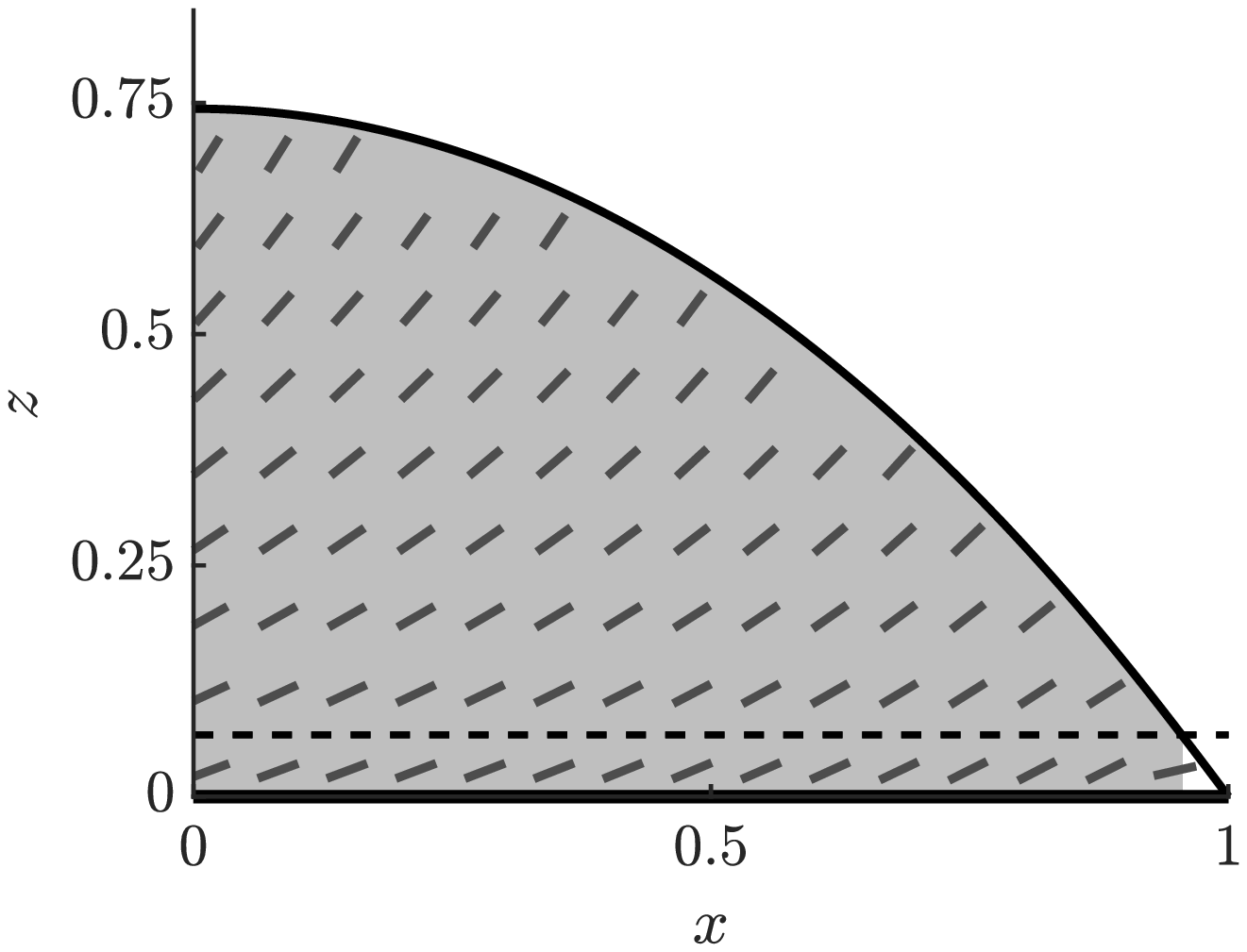} &
\includegraphics[width=0.32\linewidth]{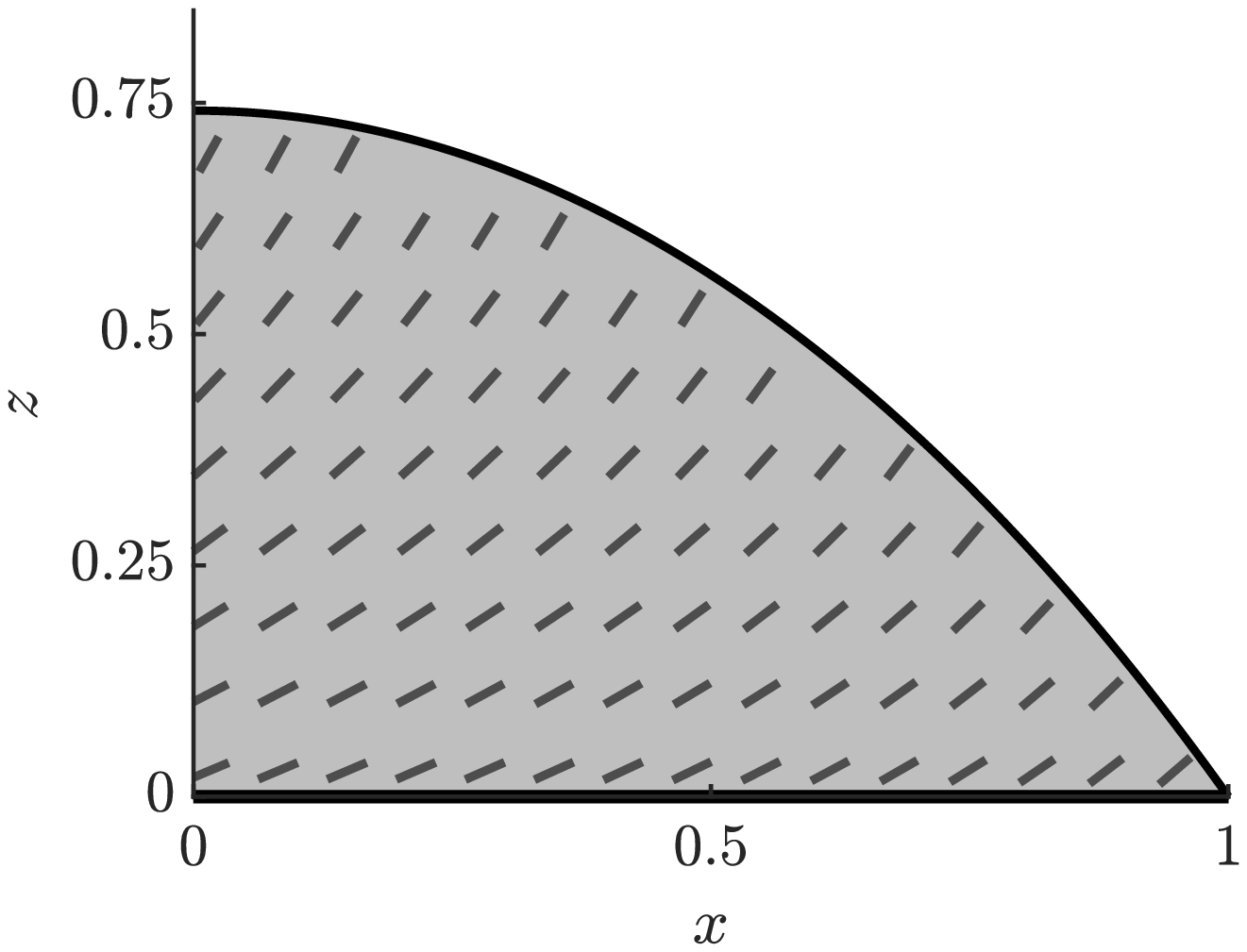} &
\includegraphics[width=0.32\linewidth]{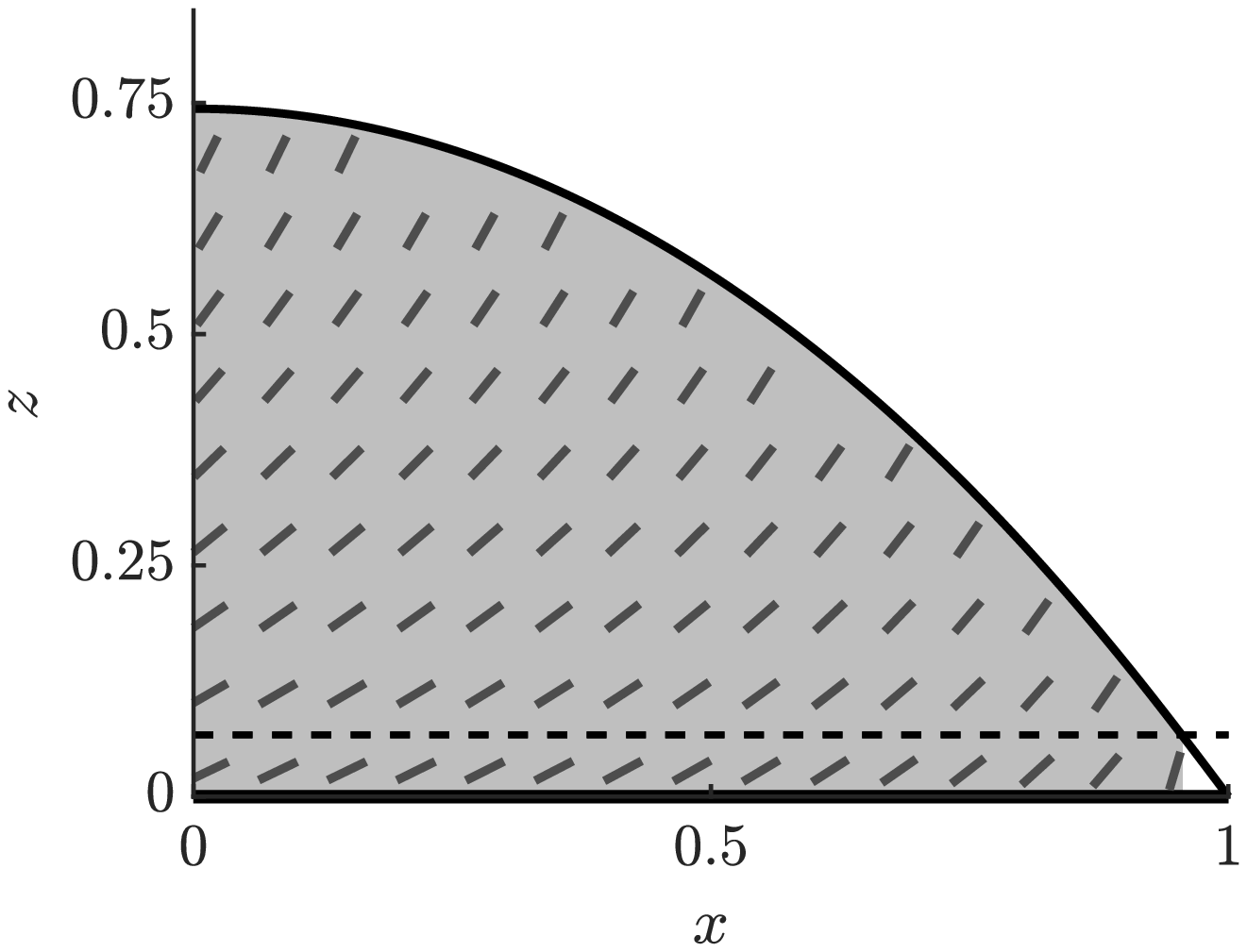} \\
(d) & (e) & (f) \\[1cm]
\includegraphics[width=0.32\linewidth]{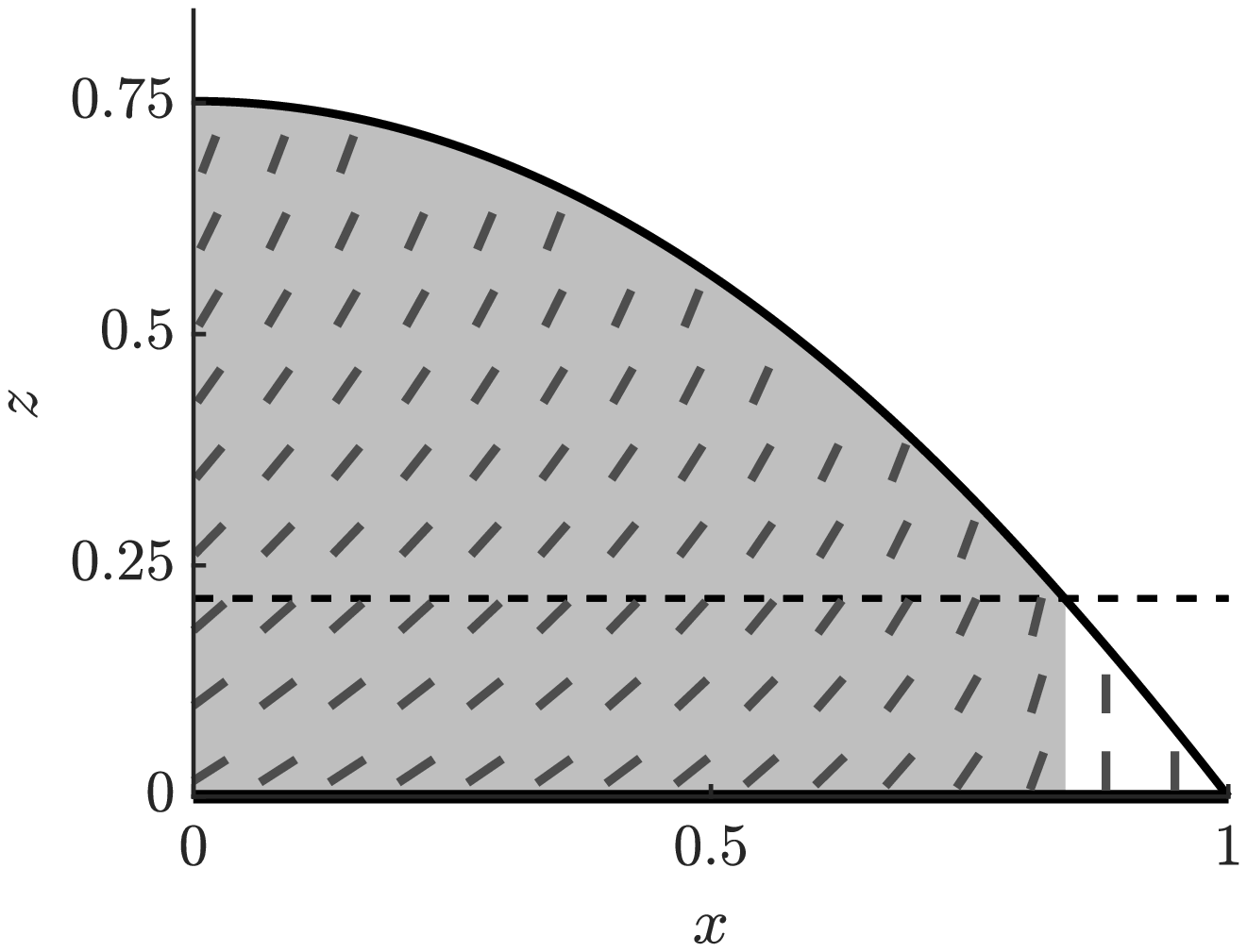} &
\includegraphics[width=0.32\linewidth]{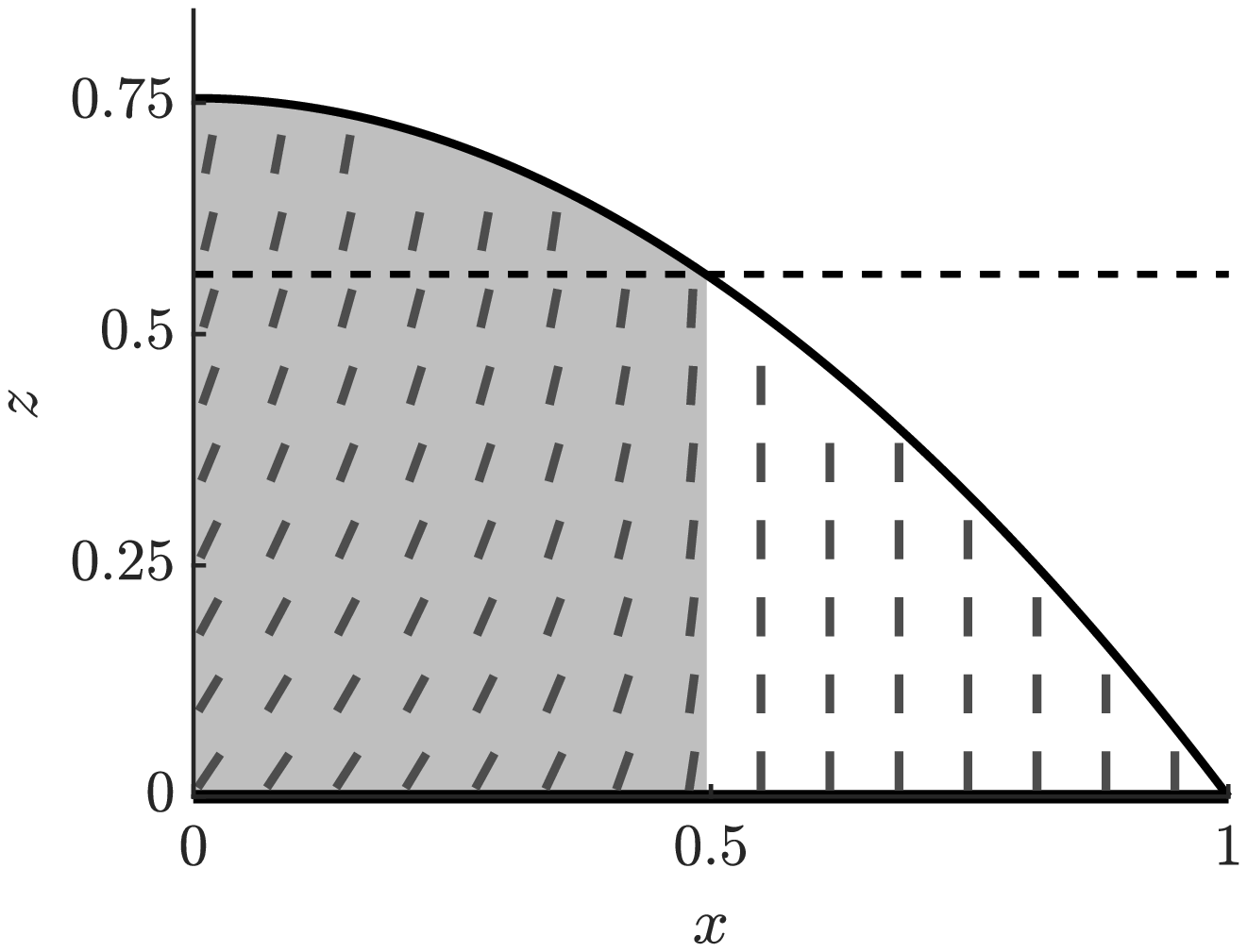} &
\includegraphics[width=0.32\linewidth]{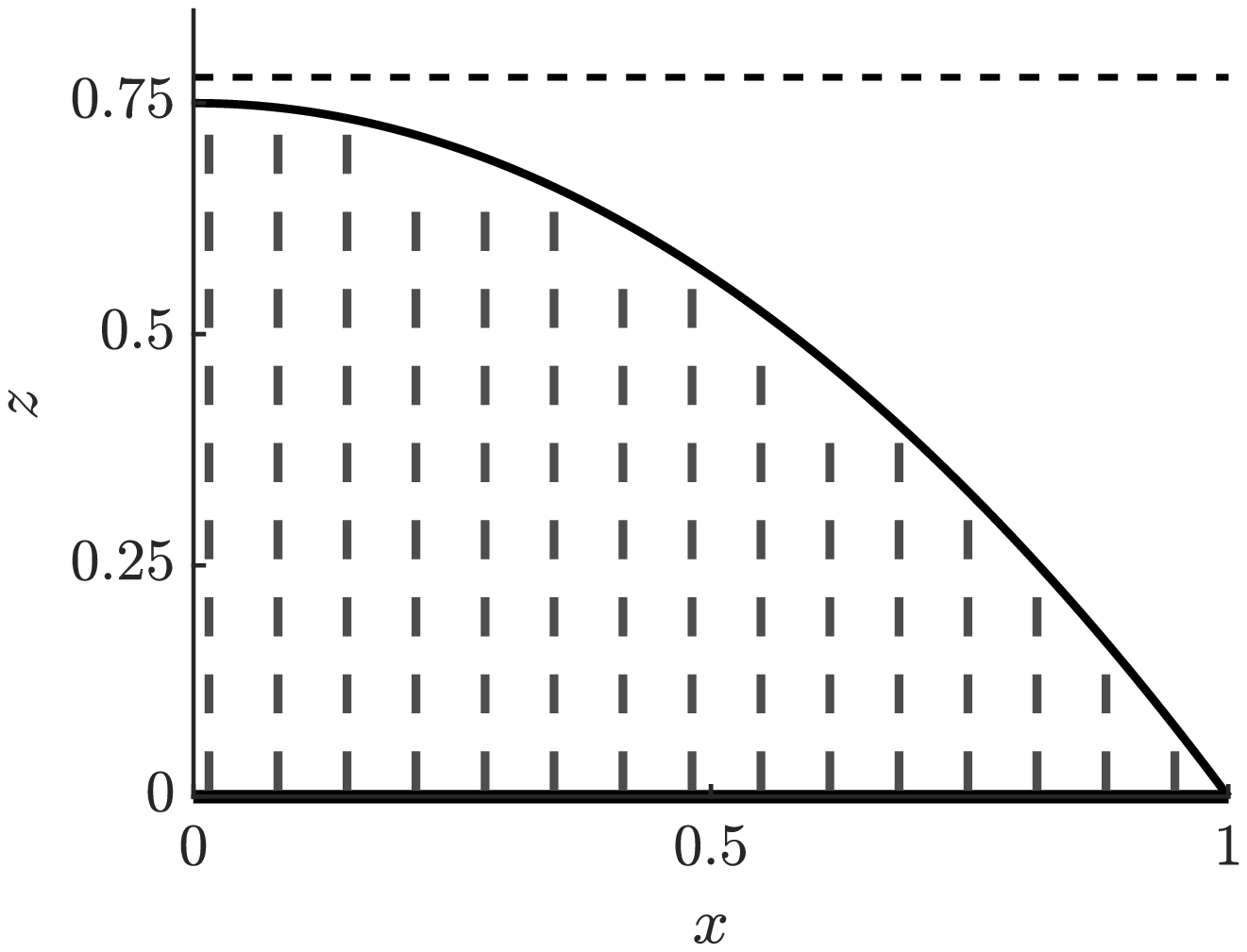} \\
(g) & (h) & (i) \\
\end{tabular}
\end{center}
\caption{
The height of the ridge $z=h$ (solid line) and the director field $\n$ (dark grey rods) for representative numerically-calculated energetically-preferred solutions when $K=1$ in the case of antagonistic anchoring with homeotropic anchoring on the gas--nematic interface and planar anchoring on the nematic--substrate interface with a constant difference between $\CNS$ and $\CGN=\CNS+5.5>0$ for the same parameter values as those used in \cref{fig:4}, \ie for (a) $\CNS=-4.5$, (b) $\CNS=-4.25$, (c) $\CNS=-3.5$, (d) $\CNS=-3$, (e) $\CNS=-2.75$, (f) $\CNS=-2.5$, (g) $\CNS=-2$, (h) $\CNS=-1.25$, and (i) $\CNS=-1$.
The critical thickness $z=h(\xc)=\vert\hc\vert$ is shown by a dashed line (note that $z=h(\xc)=\vert\hc\vert=0$ is not visible in (e)), the uniform region $\xc \le x \le 1$ in which $h\le\vert\hc\vert$ is indicated in white, and the distorted region $0 \le x < \xc$ in which $h>\vert\hc\vert$ is indicated in grey.
}
\label{fig:5}
\end{figure}

\cref{fig:4,fig:5} show
the director angles on the nematic--substrate interface $\thetaNS$ and the gas--nematic interface $\thetaGN$,
and
the height of the ridge $z=h$ and the director field $\n$,
respectively,
for representative numerically-calculated energetically-preferred solutions when $K=1$ in the case of antagonistic anchoring with homeotropic anchoring on the gas--nematic interface (\ie $\CGN>0$) and planar anchoring on the nematic--substrate interface (\ie $\CNS<0$).
In particular, \cref{fig:4,fig:5} illustrate that, depending on the values of $\CNS$ and $\CGN$, any one of the five qualitatively different types of solution can be energetically preferred.
Note that some of the solutions presented in \cref{fig:4,fig:5} are related to each other via the relations \cref{symmetrySol1,symmetrySol2,symmetrySol3}.
In particular, applying \cref{symmetrySol3} to the solutions shown in \cref{fig:4}(i), (h), (g), and (f) yields the solutions shown in \cref{fig:4}(a), (b), (c), and (d), respectively.
Also note that for the one-constant elastic constant and anchoring strengths used in \cref{fig:4,fig:5},
$h$ does not differ very much from $h=\hiso$.
Specifically, $h$ shows at most a $1\%$ difference from $h=\hiso$ for the solutions shown in \cref{fig:5}.
Although there is little change in $h$ over this range of parameter values,
there is a greater difference between $h$ and $\hiso$ for larger anchoring strengths
(see Cousins \cite{Cousinsthesis} for details).

\begin{table}[tp]
\centering
\begin{tabular}{|c|c|c|c|c|}
\hline
\quad Range of $\hc$ \quad & \quad Energetically-Preferred \quad & \quad Corresponding \quad &
\quad Director Angle \quad      & \quad Effective Refractive Index \quad \\
                           & \quad Solution \quad                & \quad Sketch        \quad &
\quad at the Contact Line \quad & \quad at the Contact Line        \quad \\
\hline
\hline
$\hc \le -3/4$ & $\Hsol$ & \cref{fig:2}(a) & $\pi/2$ & $\RIno$ \\
\hline
$-3/4 < \hc < 0$ & $\DHsol$ & \cref{fig:3}(a) & $\pi/2$ & $\RIno$ \\
\hline
$\hc = 0$ & $\Dsol$ & \cref{fig:3}(b) & $\pi/4$ & $\frac{\sqrt{2}\,\RIne\RIno}{\sqrt{{\RIne}^2+{\RIno}^2}}$ \\
\hline
$0 < \hc < 3/4$ & $\DPsol$ & \cref{fig:3}(c) & $0$ & $\RIne$ \\
\hline
$\hc \ge 3/4$ & $\Psol$ & \cref{fig:2}(b) & $0$ & $\RIne$ \\
\hline
\end{tabular}
\caption{
Classification of the energetically-preferred solutions (with corresponding sketches), the director angle at the contact line, and the effective refractive index at the contact line for antagonistic anchoring in terms of the critical thickness $\hc$.
}
\label{tab:2}
\end{table}

As mentioned in \cref{sec:distorted}, analytical progress can be made for the $\Dsol$ solution in the asymptotic limit $\CGN=-\CNS \to 0$.
In particular, in this limit, the second-order energy of $\Dsol$ solutions, given by $\DelE=3/2-\CGN^2/(4K)+\Oh(\CGN^4)$, is always less than the second-order energy of the $\Hsol$ and $\Psol$ solutions given by \cref{EHP} with $\CGN =-\CNS$, namely $\DelEH=\DelEP=3/2$, and therefore in this limit the $\Dsol$ solution is always energetically preferred to the $\Hsol$ and $\Psol$ solutions.
In fact, numerical investigations covering a wide range of parameter values including $10^{-4} \le K \le 10^4$, $\vert\CNS\vert \le 10^2$, $\vert\CGN\vert \le 10^2$ suggest that for antagonistic anchoring, distorted director solutions always have lower energy than uniform director solutions when $\vert\hc\vert < 3/4$.
As a consequence, the energetically-preferred solutions (with corresponding sketches) and the director angle at the contact line for antagonistic anchoring can be classified in terms of $\hc$ as shown in \cref{tab:2}.

\subsection{Non-antagonistic anchoring (\ie $\CNS\CGN>0$)}
\label{sec:nonantagonistic}

In contrast to the case of antagonistic anchoring described in \cref{sec:antagonistic},
for non-antagonistic anchoring the energetically-preferred solution is \emph{always} a uniform solution.
The classification of the energetically-preferred solution in terms of $\hc$ is therefore very straightforward:
for $\hc<0$ a $\Psol$ solution is preferred and for $\hc>0$ an $\Hsol$ solution is preferred.

Note that distorted director solutions are possible when $\vert\hc\vert<3/4$.
Inspection of \cref{thinEparts} with \cref{hiso} for a distorted director solution with non-antagonistic anchoring shows that $\Esurf \ge 3/2$, $\Eelas > 0$, $-\vert\CNS\vert \le \EB \le \vert\CNS\vert$, and $-\vert\CGN\vert \le \ET \le \vert\CGN\vert$, and therefore that either $\DelE > \DelEH$ or $\DelE > \DelEP$.
Hence these distorted director solutions are \emph{never} energetically preferred, and so we do not need to discuss them further.
However, we note that similar high-energy distorted director solutions in a static layer of nematic of uniform thickness bounded between two parallel substrates with non-antagonistic anchoring have been studied previously \cite{DavidsonThesis}.

\subsection{$\CNS$--$\CGN$ parameter plane}
\label{sec:parameterplane}

\begin{figure}[tp]
\begin{center}
\includegraphics[width=10cm]{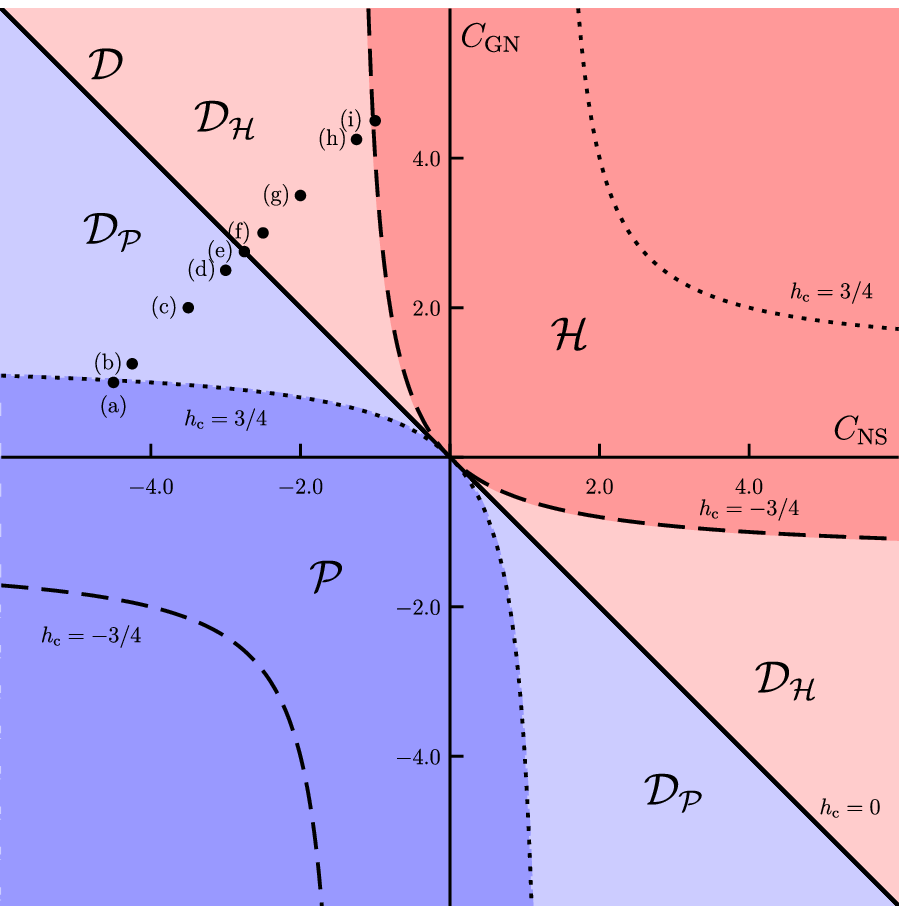}
\caption{
The $\CNS$--$\CGN$ parameter plane for the representative value $K=1$.
The regions of the parameter plane in which
an $\Hsol$  solution (shown in dark red),
a $\DHsol$ solution (shown in light red),
a $\DPsol$ solution (shown in light blue), and
a $\Psol$  solution (shown in dark blue)
is the energetically-preferred solution are indicated.
The $\Dsol$ solutions are energetically preferred along the solid line corresponding to $\hc=0$.
The dashed lines correspond to $\hc=-3/4$,
the dotted lines correspond to $\hc=3/4$, and
the points corresponding to the representative solutions shown in \cref{fig:4,fig:5}
are indicated by dots ($\bullet$) and labelled (a)--(i).
}
\label{fig:6}
\end{center}
\end{figure}

The classification obtained in \cref{sec:antagonistic,sec:nonantagonistic} can be used to determine the regions of parameter space in which the five qualitatively different types of solution are energetically preferred.

\cref{fig:6} shows the regions of the $\CNS$--$\CGN$ parameter plane for the representative value $K=1$ in which each of the five qualitatively different types of solution is the energetically-preferred solution.
In particular,
the top left and bottom right quadrants of the parameter plane
(corresponding to the case of antagonistic anchoring)
contain three curves that separate the regions of energetically-preferred solutions:
the solid line on which $\hc=0$
corresponds to the only line on which $\Dsol$ solutions are energetically preferred,
the dashed line on which $\hc=-3/4$
separates the region in which the $\Hsol$ solution is energetically preferred from the region in which the $\DHsol$ solutions are energetically preferred, and
the dotted line on which $\hc=3/4$
separates the region in which the $\Psol$ solution is energetically preferred from the region in which the $\DPsol$ solutions are energetically preferred.
The points corresponding to the representative solutions shown in \cref{fig:4,fig:5}
are indicated by dots ($\bullet$) and labelled (a)--(i).
Using \cref{hc}, the curves on which $\hc=-3/4$ and $\hc=3/4$ can be expressed in terms of the parameters $K$, $\CNS$, and $\CGN$ by
\begin{align}\label{specialHC1}
\CGN &= -\dfrac{4K\CNS}{3\CNS+4K}
\end{align}
and
\begin{align}\label{specialHC2}
\CGN &=  \dfrac{4K\CNS}{3\CNS-4K},
\end{align}
respectively.
In the limit $\CNS \to \infty$ the curves \cref{specialHC1,specialHC2}
approach $\CGN=-4K/3$ and $\CGN=4K/3$, respectively, from above, and
in the limit $\CNS \to -\infty$ they
approach the same values from below.
In particular, \cref{fig:6} shows that for $\CNS<-4K/3$ and $\CGN<-4K/3$ the $\Hsol$ solution is never energetically preferred and, similarly, for $\CNS>4K/3$ and $\CGN>4K/3$ the $\Psol$ solution is never energetically preferred.
This conclusion may be important for industrial applications involving pinned nematic ridges, where uniform homeotropic or planar director alignment (\ie the $\Hsol$ or $\Psol$ solutions) could be eliminated by selecting the anchoring strengths of the nematic--substrate ($i={\rm NS}$) and/or the gas--nematic ($i={\rm GN}$) interfaces to satisfy $\vert c_i\vert > 4K/3$.

As $K$ is varied, the curves on which $\hc=-3/4$ and $\hc=3/4$ given by \cref{specialHC1,specialHC2} vary; however, the qualitative behaviour shown in \cref{fig:6} remains the same.
In the limit $K \to 0$, the curves on which $\hc=\pm 3/4$ approach the axes $\CNS=0$ and $\CGN=0$, indicating that $\DHsol$ and $\DPsol$ solutions are energetically preferred for all situations with antagonistic anchoring when elastic effects are weak.
Similarly, in the limit $K \to \infty$, the curves on which $\hc=\pm 3/4$ approach the diagonal straight line $\CNS=-\CGN$, indicating that the $\Hsol$ and $\Psol$ solutions are energetically preferred for all situations with antagonistic anchoring when elastic effects are strong.

\subsection{The effective refractive index of the ridge}
\label{sec:neff}

\begin{figure}[tp]
\begin{center}
\includegraphics[width=0.5\linewidth]{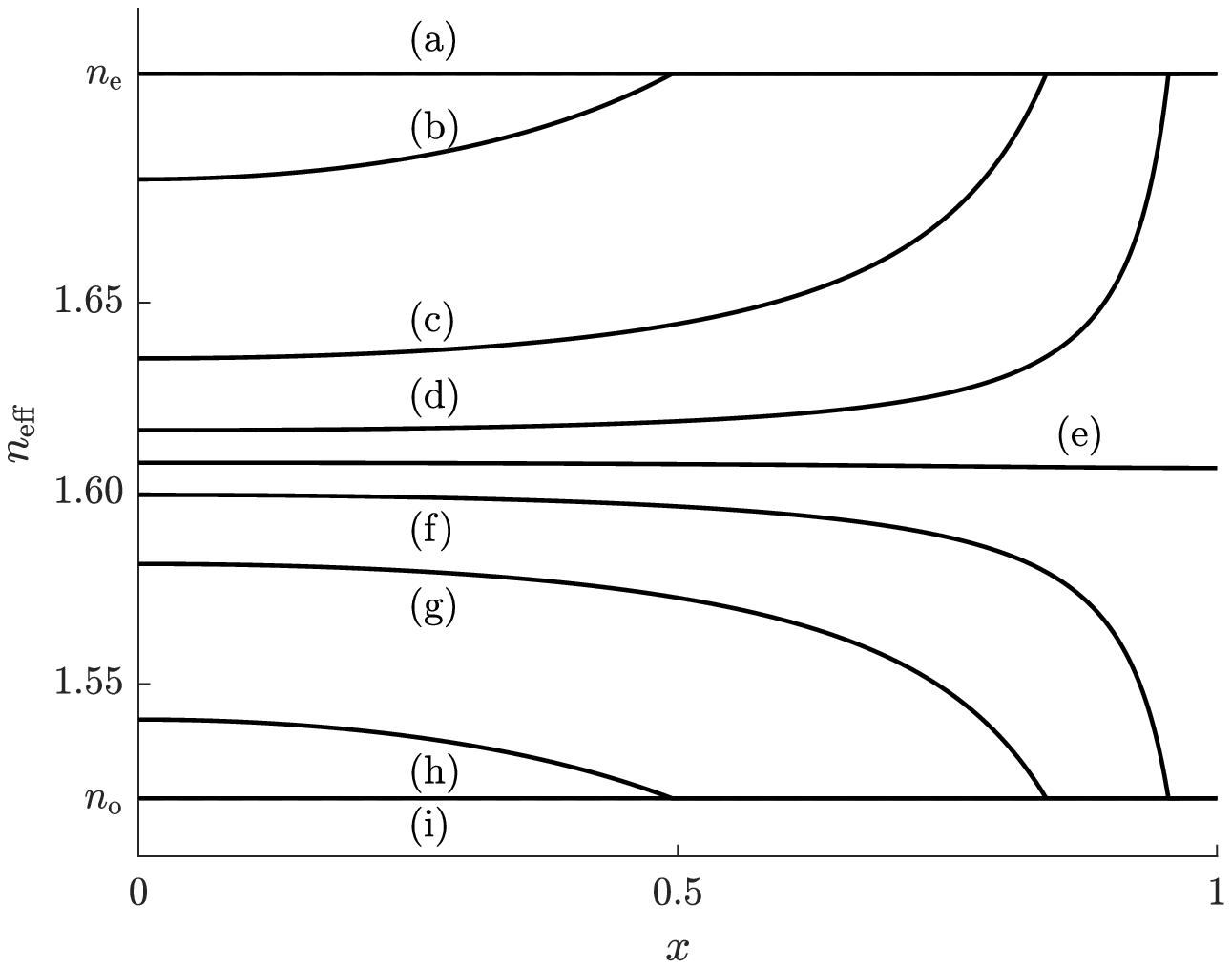}
\end{center}
\caption{
The effective refractive index $\neff$ given by (\ref{neff3}) for the representative solutions labelled (a)--(i) shown in \cref{fig:4,fig:5,fig:6} for $\RIne=1.71$ and $\RIno=1.52$.
}
\label{fig:7}
\end{figure}

In order to compare the theoretical predictions with the results of the physical experiments described in \cref{sec:experiments}, we now consider an experimentally measurable quantity, namely the effective refractive index of the ridge for $x$-polarised light transmitted through the ridge in the $z$-direction, denoted by $\neff=\neff(x)$ and defined by
\begin{align}\label{neff3}
\neff &= \dfrac{1}{h} \int^{h}_0 \dfrac{\RIne\RIno}{\sqrt{{\RIne}^2\sin^2 \theta+{\RIno}^2\cos^2 \theta}} \, \dd z,
\end{align}
where $\RIne$ and $\RIno$ are the extraordinary and ordinary refractive indices of the nematic [65].
We note that \cref{neff3} can be expressed in terms of the director angles $\thetaNS$ and $\thetaGN$ as
\begin{align}\label{neff2}
\neff &=
\begin{cases}
\qquad\qquad\quad \RIne & \text{\: for \quad $\thetaNS=\thetaGN\equiv0$,} \\
\dfrac{F(\thetaGN,k) - F(\thetaNS,k)}{\thetaGN-\thetaNS}\,\RIne & \text{\: for \quad $\thetaNS\neq\thetaGN$,} \\
\qquad\qquad\quad \RIno & \text{\: for \quad $\thetaNS=\thetaGN\equiv\pi/2$,}
\end{cases}
\end{align}
where $F(\theta_i,k)$ with $i={\rm NS}$ or $i={\rm GN}$ are elliptic integrals of the first kind, defined by
\begin{align}
F(\theta_i,k) &= \int^{\theta_i}_0 \dfrac{\dd u}{\sqrt{1-k\sin^2u}},
\end{align}
and $k=1-(\RIne/\RIno)^2$.
For the nematic 5CB used in the experiments discussed in \cref{sec:experiments},
$\RIne=1.71$, $\RIno=1.52$, and hence $k=-0.27$ \cite{PHYSICALPROPERTIES2001}.

\cref{fig:7} shows the effective refractive index $\neff$ for the representative solutions labelled (a)--(i) in \cref{fig:4,fig:5,fig:6} for $\RIne=1.71$ and $\RIno=1.52$.
In particular,
\cref{fig:7} shows that for $\Psol$ and $\Hsol$ solutions,
illustrated by the curves labelled (a) and (i),
the effective refractive index takes the constant values $\neff=\RIne$ and $\neff=\RIno$, respectively.
For $\DPsol$ solutions,
labelled (b)--(d),
$\neff$ increases monotonically in the distorted region to a constant value of $\neff=\RIne$ in the uniform region.
For $\Dsol$ solutions,
labelled (e),
$\neff$ is constant.
The value of this constant can be determined by recalling that
for a $\Dsol$ solution $\thetaNS = \thetaGN = \pi/4$ at $x=1$
and
taking the limits $\thetaNS \to {\pi/4}^{\pm}$ and $\thetaGN \to {\pi/4}^{\mp}$ in \cref{neff2} to obtain
\begin{equation}\label{neffvalue}
\neff = \frac{\sqrt{2}\,\RIne\RIno}{\sqrt{{\RIne}^2+{\RIno}^2}}.
\end{equation}
In particular,
(\ref{neffvalue}) with $\RIne=1.71$ and $\RIno=1.52$ gives the value $\neff=1.61$ shown in \cref{fig:7}.
For $\DHsol$ solutions,
labelled (f)--(h),
$\neff$ decreases monotonically in the distorted region to a constant value of $\neff=\RIno$ in the uniform region.
A classification of the effective refractive index at the contact line for antagonistic anchoring in terms of $\hc$ is included in \cref{tab:2}.

\section{Experimental investigation of a ridge of 5CB}
\label{sec:experiments}

Physical experiments were performed for a pinned static ridge of the nematic 5CB, the results of which support the theoretical predictions presented in \cref{sec:ridge,sec:uniform,sec:distorted,sec:preferred}.

\subsection{Experimental setup}
\label{sec:setup}

\begin{figure}[tp]
\begin{center}
\setlength\tabcolsep{1pt}
\begin{tabular}{ccc}
\quad \includegraphics[height=6cm]{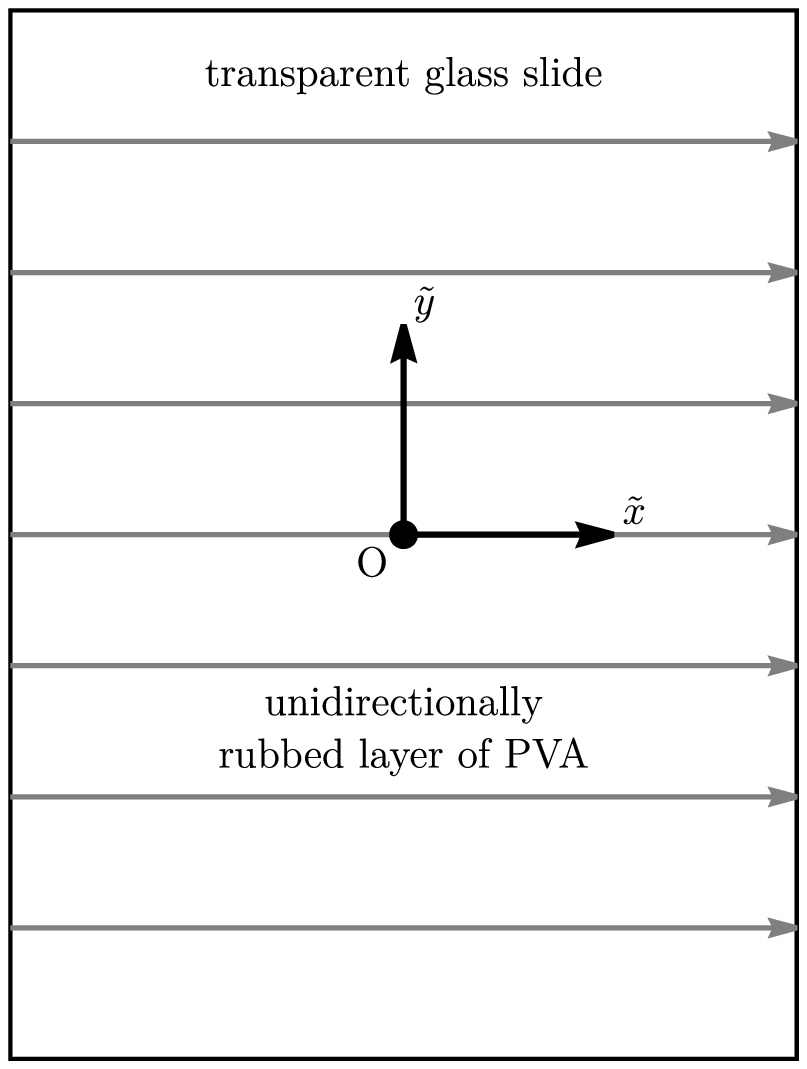} \quad &
\quad \includegraphics[height=6cm]{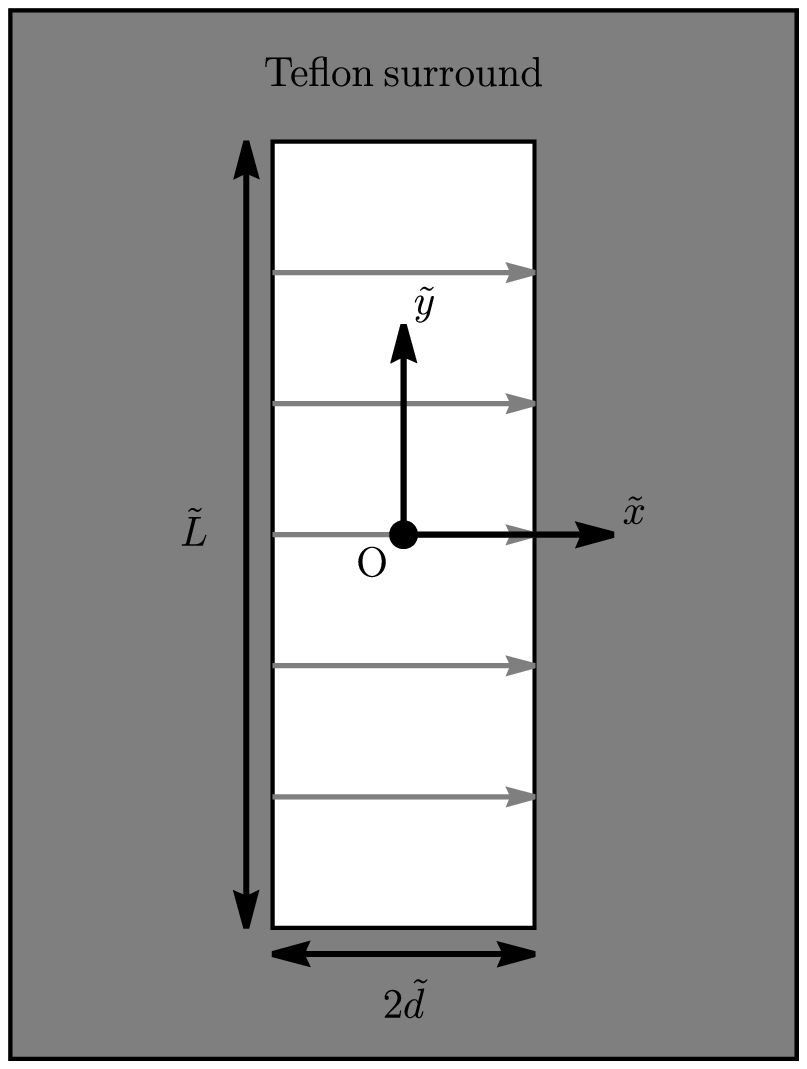} \quad &
\quad \includegraphics[height=6cm]{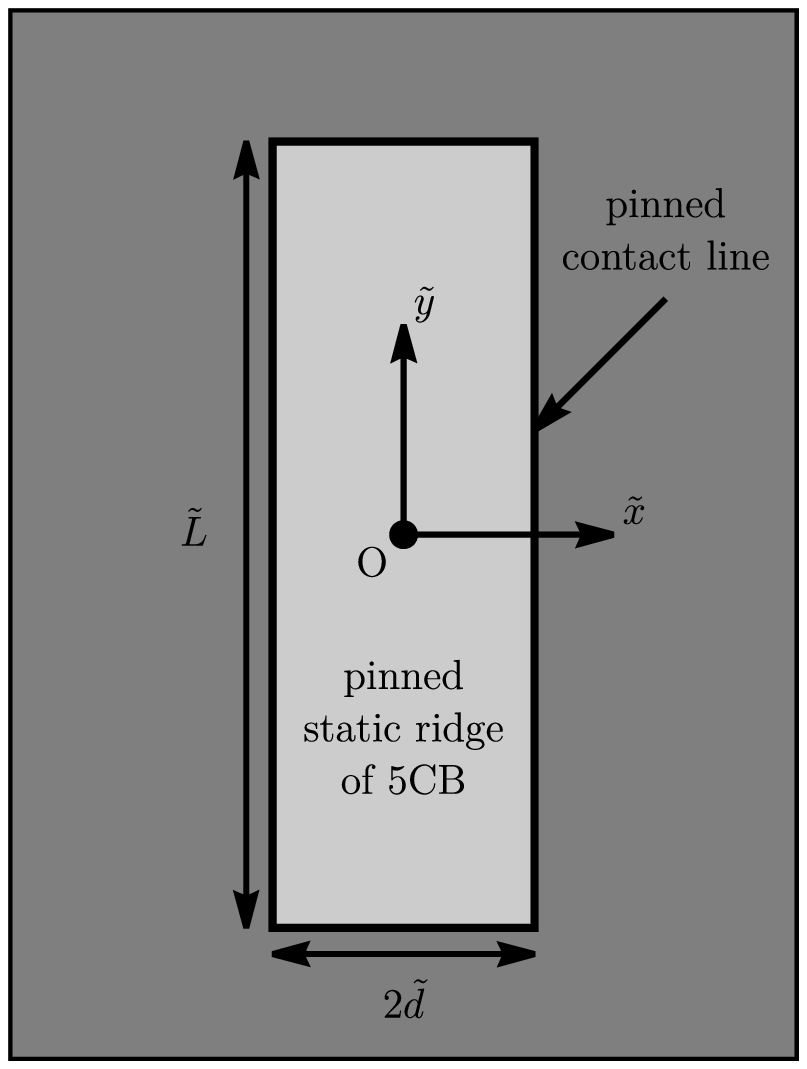} \quad \\
(a) & (b) & (c)
\end{tabular}
\end{center}
\caption{
Schematic diagrams showing a plan view of
(a) a transparent glass slide coated with a layer of PVA unidirectionally rubbed in the $\tx$-direction (shown by the grey arrows),
(b) a unidirectionally rubbed PVA--glass substrate (the white rectangular region with the grey arrows) of length $\tL$ in the $\ty$-direction and width $2\td$ in the $\tx$-direction with a Teflon surround (the dark grey region), and
(c) a pinned static ridge of 5CB (the light grey rectangle).
The Cartesian coordinates $\tx$ and $\ty$ (with the $\tz$-axis out of the page) are also indicated.
}
\label{fig:8}
\end{figure}

A transparent glass slide was coated with a layer of the polymer poly(vinyl alcohol) (PVA) (CAS 9002-89-5, molecular weight 9000--10000, Sigma-Aldrich/Merck KGaA, Darmstadt, Germany), which was unidirectionally rubbed in the $\tx$-direction, as shown in \cref{fig:8}(a), creating a homogeneous planar director alignment in the $\tx$-direction with a low pretilt \cite{Yue2012}.
The glass slide coated with the layer of PVA was then partially coated, using a masking process, with an amorphous fluorinated copolymer Teflon AF (CAS 37626-13-4, Sigma-Aldrich/Merck KGaA, Darmstadt, Germany), creating a unidirectionally rubbed PVA--glass substrate of length $\tL=6\times10^{-2}\,$m in the $\ty$-direction and width $2\td=1.2\times10^{-3}\,$m in the $\tx$-direction with a Teflon surround, as shown in \cref{fig:8}(b).
The experiments were conducted in an atmosphere of air at an ambient temperature of $(22 \pm 1.5)^\circ$C.

A sample of the nematic 5CB (CAS 40817-08-1, Sigma-Aldrich/Merck KGaA, Darmstadt, Germany) with an estimated volume of $1\times10^{-9}\,$m$^{3}$ was then prepared in a solution (10$\%$ by weight) with toluene (CAS 108-88-3, Sigma-Aldrich/Merck KGaA, Darmstadt, Germany) and deposited on to the substrate.
A pinned static ridge of 5CB, as shown in \cref{fig:8}(c), was then created by thermally annealing the deposited 5CB--toluene solution at 50\,$^\circ$C.
After the annealing process was finished, the nematic-to-isotropic phase transition of the deposited 5CB was observed to occur at $35.2\,^{\circ}{\rm C}$ \cite{Gray1973,IntroLC1997}, indicating complete evaporation of the toluene.
The Teflon surround provided pinning of the contact line, so that the ridge had length $\tL$ in the $\ty$-direction and width $2\td$ in the $\tx$-direction with contact lines along the length of the ridge at $\tx=\pm\td=\pm600\,\mu$m and centreline at $\tx=0$.
The entire process was then repeated for a second sample of 5CB with an estimated volume of $3\times10^{-9}\,$m$^{3}$ to create a second pinned static ridge of 5CB with a larger volume.
As we shall see shortly in \cref{sec:inf}, the ridge with an estimated volume of $1\times10^{-9}\,$m$^{3}$, hereafter referred to as \emph{the small ridge}, corresponds to a smaller aspect ratio $\epsilon$ than the ridge with an estimated volume of $3\times10^{-9}\,$m$^{3}$, hereafter referred to as \emph{the large ridge}.

The preferred director alignment on the nematic--substrate interface has previously been determined to be planar in experiments involving an interface between glass coated with a layer of PVA and 5CB, with a well-characterised anchoring strength in the range $\vert\tCNS\vert=10^{-4}$--$10^{-3}\,{\rm N m}^{-1}$ \cite{Yokoyama1988,Feller1991,Stoenescu1996,Parshin2020}.
The preferred director alignment on the gas--nematic interface has previously been determined to be homeotropic \cite{Ohzono2012}; this was confirmed using a separate experiment on a free-standing film of 5CB.
To the best of our knowledge, there have been no measurements of the anchoring strength of an interface between 5CB and air; however, we anticipate that this anchoring strength will be significantly weaker than the anchoring strength of the nematic--substrate interface \cite{SONINBOOK1995}.
In particular, we expect that $-\tCNS >\tCGN > 0$, and therefore from the results summarised in \cref{tab:1} we expect that $\hc > 0$, and hence from the classification summarised in \cref{tab:2} that either a $\Psol$ solution or a $\DPsol$ solution will occur.

\begin{figure}[tp]
\begin{center}
\includegraphics[width=0.4\linewidth]{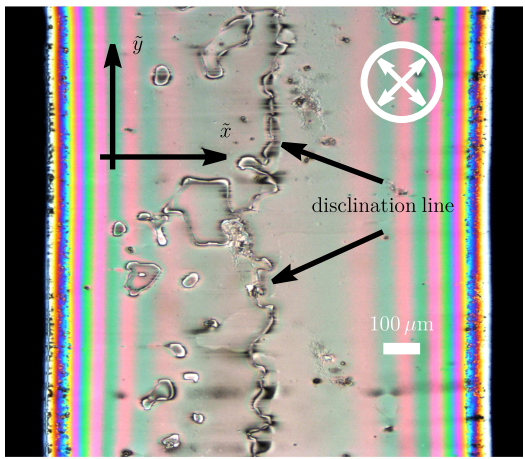}
\end{center}
\vspace{-1cm}
\caption{
Polarising optical micrograph of the small ridge of 5CB.
The orientations of the polariser and the analyser are shown by the two crossed white arrows enclosed within the white circle.
The Cartesian coordinates $\tx$ and $\ty$ (with the $\tz$-axis out of the page) and a $100\,\mu$m scale bar are also indicated.
}
\label{fig:9}
\end{figure}

\cref{fig:9} shows a polarising optical micrograph of the small ridge, taken using a polarising optical microscope with broadband polychromatic illumination.
The orientations of the polariser and the analyser are shown by the two crossed white arrows enclosed within the white circle in \cref{fig:9}.
The dark regions on the left-hand and right-hand sides of \cref{fig:9} correspond to the optically-isotropic Teflon surround, whereas the optically-anisotropic nematic produces transmitted light with a colour that depends on the height of the ridge $\tilh$, the birefringence of 5CB, and the director configuration.
In particular, \cref{fig:9} shows two regions of opposing director distortion either side of a disclination line (which is not straight as a result of the influence of unavoidable inhomogeneities in the system) near to the centreline of the ridge.
In particular, as mentioned in \cref{sec:discontinuous},
the presence of a disclination line 
motivates allowing for the occurrence of a discontinuity in $\theta$ at the centreline of the ridge.

\subsection{Interferometry measurements of a ridge of 5CB}
\label{sec:inf}

In order to determine the structure of a pinned static ridge of 5CB close to the contact lines, the right-hand edges (specifically, the region $400\,\mu$m$\,\le \tx \le 600 \, \mu$m) of the two ridges described in \cref{sec:setup} were analysed using the displacement of tilt fringes in a Mach--Zehnder interferometer \cite{Trabi2008} illuminated by collimated monochromatic He--Ne laser light of wavelength $632.8\,$nm incident on the ridges in the $\tz$-direction.
The laser beam in the interferometer was expanded after the spatial light filter, and direct in situ microscopic imaging of the sample was permitted using a $4f$ relay lens.
First, the sample was imaged in transmission mode by blocking the laser beam in one of the arms of the interferometer.
This provided an accurate determination of the position of the edge of the ridge (with an error in the $\tx$-direction of $\pm 4\,\mu$m), and hence there was little (if any) overlap of the 5CB and the Teflon surround.
Then the tilt fringe interference pattern was created using light beams from both arms of the interferometer with the sample in situ.
The fringe pattern produced by the interferometer depends on the orientation of the linear polarisation of the incident laser light.
An analysis of these fringe patterns enables the height of the ridge $\tilh=\tilh(\tx)$ and the effective refractive index $\neff=\neff(\tx)$ to be calculated \cite{YehBook2005}.
In order to perform this calculation we assume that the planar director alignment in the $\tx$-direction on the substrate ensured that the director did not vary in the $\ty$-direction, allowing $\tilh$ to be obtained from the fringe pattern when the polarisation of the incident laser light was in $\ty$-direction.
A comparison of the fringe patterns when the polarisation was in the $\tx$-direction and $\ty$-direction was then used to determine $\neff$ and an estimate of the associated error \cite{YehBook2005}.
The uncertainties in the values obtained for $\neff$ reflect the precision with which the centres of the fringes can be measured, which includes uncertainties due to non-uniformities in the intensity of the fringe pattern, as well as uncertainties in the horizontal resolution which affects the precision with which the edge of the ridge can be located.

\begin{figure}[tp]
\begin{center}
\includegraphics[width=0.5\linewidth]{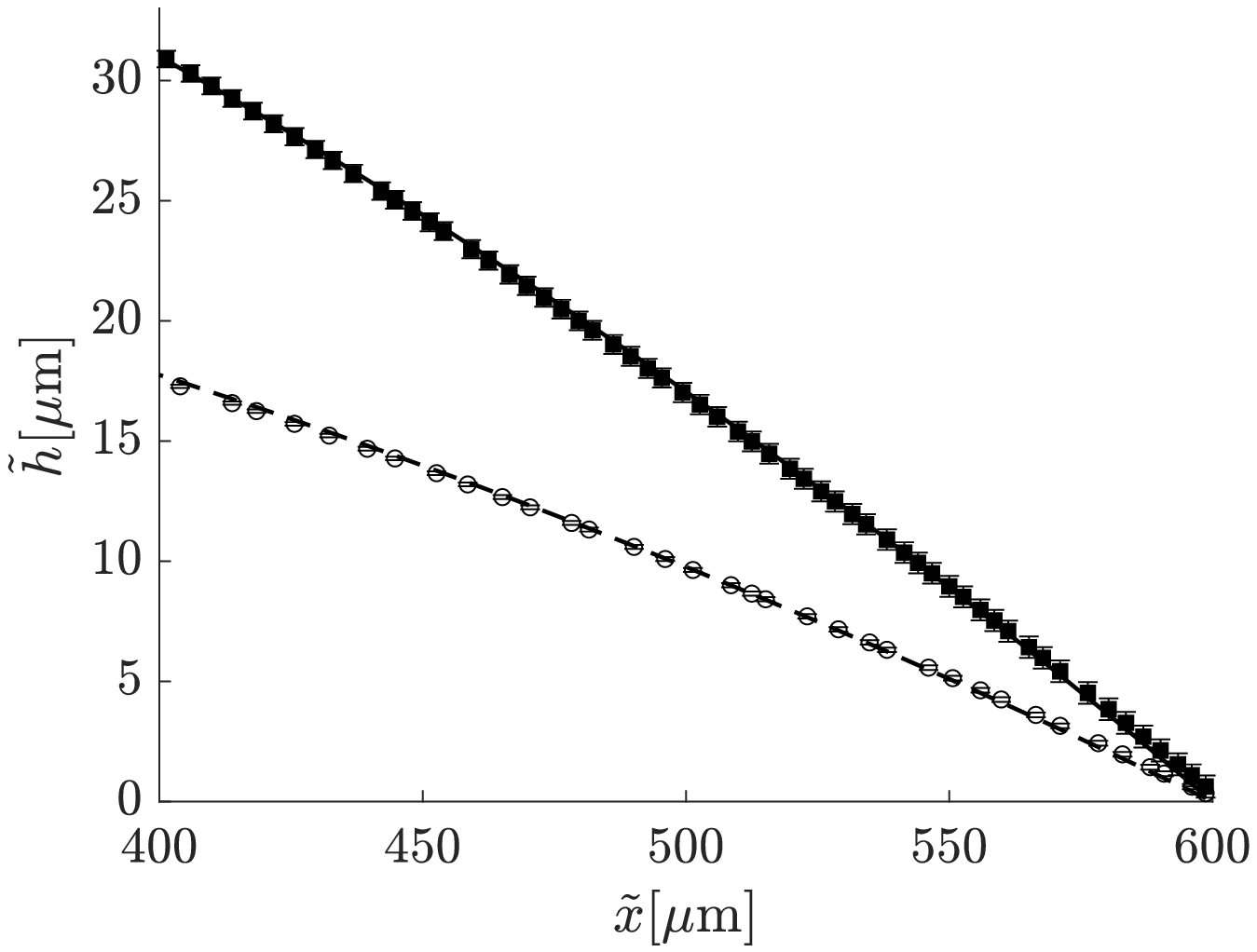}
\end{center}
\caption{
The experimental values of the height of the ridge $\tilh$ plotted as a function of $\tx$ for the right-hand edge (specifically, the region $400\,\mu$m$\,\le \tx \le 600 \,\mu$m) of the small ridge (shown by open circles) and the large ridge (shown by closed squares).
The fitted theoretical predictions for a pinned static isotropic ridge $\thiso=3\epsilon(\td^2-{\tx}^2)/(4\td)$ as a function of $\tx$ with $\epsilon=0.071$ (dashed line) and $\epsilon=0.124$ (solid line), and $2\td=1.2\times10^{-3}\,$m are also plotted.
}
\label{fig:10}
\end{figure}

\cref{fig:10} shows the experimental values of the height of the ridge $\tilh$ plotted as a function of $\tx$ for the right-hand edge of the small ridge (shown by open circles) and the large ridge (shown by closed squares).
The experimental values of $\tilh$ for each ridge show a quadratic profile that is in qualitative agreement with the corresponding theoretical prediction for a pinned static isotropic ridge given by \cref{hiso} with \cref{Scale}, namely $\thiso=3\epsilon(\td^2-{\tx}^2)/(4\td)$.
To make a quantitative comparison between the experimental values and the theoretical prediction, the values of the aspect ratio $\epsilon$ given by \cref{ep} must be determined for each ridge.
We determined $\epsilon$ by using a least-squares fit between $\thiso$ for $2\td=1.2\times10^{-3}\,$m and the experimental values of $\tilh$ using $\epsilon$ as a fitting parameter, which gave the values $\epsilon=0.071$ for the small ridge and $\epsilon=0.124$ for the large ridge.
\cref{fig:10} shows that excellent agreement is found between the experimental values and the fitted theoretical predictions for a pinned static isotropic ridge.

\begin{figure}[tp]
\begin{center}
\includegraphics[width=0.5\linewidth]{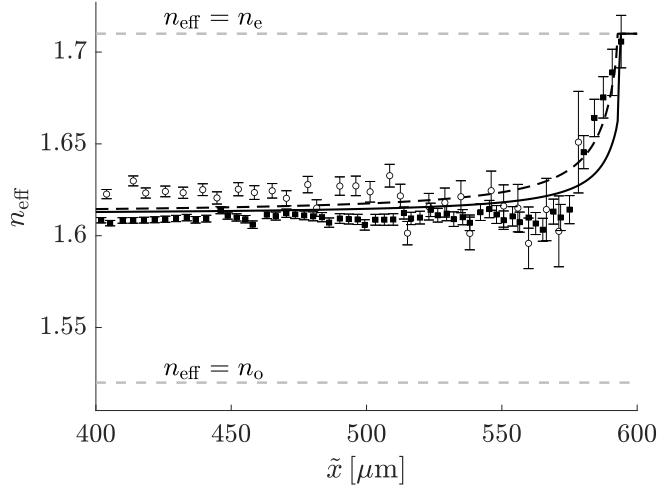}
\end{center}
\caption{
The experimental values of the effective refractive index $\neff$ plotted as a function of $\tx$ for the right-hand edge (specifically, the region $400\,\mu$m$\,\le \tx \le 600 \,\mu$m) of the small ridge (open circles) and the large ridge (closed squares).
The theoretical predictions for $\neff$ as a function of $\tx$ with $\epsilon= 0.071$ (dashed line) and $\epsilon=0.124$ (solid line), $\tCNS=-5.0\times10^{-4}\,{\rm N m}^{-1}$, $\tCGN=9.80\times10^{-6}\,{\rm N m}^{-1}$, $\tK=7.2\times10^{-12}\,$N, $\RIne=1.71$ and $\RIno=1.52$ are also plotted.
The extraordinary and ordinary refractive indices of the nematic $\neff=\RIne$ (dashed grey line) and $\neff=\RIno$ (dashed grey line), respectively, are also plotted.
For clarity, the error of $\pm 4\,\mu$m in the $\tx$-direction is omitted from the above plots.
}
\label{fig:11}
\end{figure}

\cref{fig:11} shows the experimental values of the effective refractive index $\neff$ plotted as a function of $\tx$ for the right-hand edge of the small ridge (shown by open circles) and the large ridge (shown by closed squares).
The experimental values of $\neff$ for each ridge show that close to the contact lines $\neff$ increases as $\tx$ increases for both ridges, which indicates that the director approaches a planar alignment as the contact line is approached.
Specifically, the experimental values of $\neff$ show that the homeotropic anchoring at the gas--nematic interface is broken close to the contact lines of both ridges, and thus that the anchoring strength of the nematic--substrate interface is greater than that of the gas--nematic interface.
Away from the contact line (specifically, in the region $\tx \le 550\,\mu$m), $\neff$ is approximately constant with the average values $\neff=1.62$ and $\neff=1.61$ for the small and the large ridge, respectively.
Note that for a distorted director solution with a $\pi/2$ rotation of the director between $\tz=0$ and $\tz=\tilh$, \ie $\thetaNS=0$, $\thetaGN=\pi/2$ and hence $\theta=\pi\tz/(2\tilh)$, equation \cref{neff2} reduces to $\neff=2\RIno F(\pi/2,k)/\pi=1.61$.
The experimental values of $\neff$ shown in \cref{fig:11} therefore indicate that the director is distorted away from the contact line and tends towards a planar alignment as the contact line is approached, in qualitative agreement with the behaviour of the $\DPsol$ solution predicted theoretically and shown by the curves labelled (b)--(d) in \cref{fig:7}.

\subsection{Anchoring strength of the gas--nematic interface}
\label{sec:anchoringstrength}

Before making any further quantitative comparisons between the experimental results, shown in \cref{fig:11}, and the theoretical predictions for $\neff$, it is useful to review which of the parameter values are known and which are unknown.
The anchoring strength of the nematic--substrate interface for 5CB, the elastic constants of 5CB, and the extraordinary and ordinary refractive indices of 5CB are well characterised, and hence values are readily available for these parameters.
However, as discussed in \cref{sec:governing}, the anchoring strength of the gas--nematic interface for 5CB is unknown.
We will therefore compare the experimental values and the theoretical predictions for $\neff$ to give a first estimate of $\tCGN$ for this system.
Specifically, we use a least-squares fit between the experimental values and theoretical predictions to estimate $\tCGN$ using
$\vert\tCNS\vert=10^{-4}$--$10^{-3}\,{\rm N m}^{-1}$ \cite{Yokoyama1988,Feller1991,Stoenescu1996,Parshin2020},
values of $\tK$ lying between
the splay elastic constant $\tK_1=6.2\times10^{-12}\,$N
and
the bend elastic constant $\tK_3=8.2\times10^{-12}\,$N
for 5CB \cite{ISBOOK2004},
$\epsilon=0.071$ for the small ridge, and
$\epsilon=0.124$ for the large ridge,
which gives an estimate of the anchoring strength of an interface between air and the nematic 5CB to be
\begin{align}
\tCGN=(9.80\pm1.12)\times10^{-6}\,{\rm N m}^{-1}.
\end{align}
Note that the uncertainty of $\pm1.12\times10^{-6}\,{\rm Nm}^{-1}$ is calculated using the upper and lower bounds on $\tCNS$ and $\tK$.
The spatial variation in the $\tx$-direction of the optical path length that is experienced by light travelling through the ridge in the $\tz$-direction leads to lensing effects.
The calculated lensing effects in the experiments are of a similar magnitude to those found in similar experiments using interferometry and, as suggested in Ottevaere and Thienpont \cite{Ottevaere2005}, we do not expect that lensing effects introduce any significant error.
\cref{fig:11} shows good agreement between the prediction of the theory for $\neff$ with $\epsilon=0.071$ (dashed line) and $\epsilon=0.124$ (solid line), $\tCNS=-5.0\times10^{-4}\,{\rm N m}^{-1}$ and $\tCGN=9.80\times10^{-6}\,{\rm N m}^{-1}$ and the experimental values of $\neff$ for the small and the large ridge, respectively.
This estimate of the anchoring strength of the gas--nematic interface is consistent with the experimental measurements of the anchoring strength of the gas--nematic interface for ZLI 2860 \cite{FreeSurfaceAnchoring1997} and MBBA \cite{Chiarelli1983} mentioned in \cref{sec:governing}.

\section{Conclusions}

In the present work, we performed a theoretical investigation of weak-anchoring effects in a thin two-dimensional pinned static ridge of nematic liquid crystal resting on a flat solid substrate in an atmosphere of passive gas.
Specifically, we solved a reduced version of the general system of governing equations recently derived by Cousins \etal \cite{Cousins2022} valid for a symmetric thin ridge under the one-constant approximation of the Frank--Oseen bulk elastic energy with pinned contact lines to determine the shape of the ridge and the behaviour of the director within it.
Numerical investigations covering a wide range of parameter values indicated that the energetically-preferred solutions can be classified in terms of the critical thickness $\hc$ given by \cref{hc}.
Specifically,
for antagonistic anchoring the energetically-preferred solution is
an $\Hsol$ solution, a $\DHsol$ solution, a $\Dsol$ solution, a $\DPsol$ solution, and a $\Psol$ solution for
$\hc \le -3/4$, $-3/4 < \hc < 0$, $\hc=0$, $0 < \hc < 3/4$, and $\hc \ge 3/4$, respectively, while
for non-antagonistic anchoring the energetically-preferred solution is
a $\Psol$ solution for $\hc <0$ and an $\Hsol$ solution for $\hc > 0$.
In particular,
the theoretical results suggest that anchoring breaking occurs close to the contact lines,
validating the analysis of \cite{Cousins2022} in the present situation.
The theoretical predictions were supported by the results of physical experiments for a ridge of the nematic 5CB.
In particular, these experiments showed that the homeotropic anchoring at the gas--nematic interface is broken close to the contact lines by the stronger planar anchoring at the nematic--substrate interface.
A comparison between the experimental values of and the theoretical predictions for $\neff$ gave a first estimate of the anchoring strength of an interface between air and 5CB to be $(9.80\pm1.12)\times10^{-6}\,{\rm N m}^{-1}$ at a temperature of $(22\pm1.5)^\circ$C.

\section*{Conflicts of Interest}

The authors have no competing interests to declare.

\section*{Data Accessibility Statement}

The theoretical data used to produce \cref{fig:4,fig:5,fig:6} may be generated using the numerical procedure for solving the system \cref{thinNS,thinGN,thinYL,thinBC,thinsym,thinArea} described in \cref{app:num}.
The experimental data that support the findings of this study are available upon request from Prof.\ Carl Brown (carl.brown@ntu.ac.uk) at Nottingham Trent University.

\section*{Acknowledgements}

The theoretical work was supported by the United Kingdom Engineering and Physical Sciences Research Council (EPSRC), the University of Strathclyde, the University of Glasgow, and Merck KGaA via EPSRC research grants EP/P51066X/1 and EP/T012501/2.
The experimental work, which was conducted at Nottingham Trent University, was supported by the EPSRC via EPSRC research grant EP/T012986/1.

\appendix

\section{Numerical procedure for solving the system \cref{thinNS,thinGN,thinYL,thinBC,thinsym,thinArea}}
\label{app:num}

The numerical solutions presented in the current work were obtained with the programming and numerical computing platform MATLAB \cite{MATLAB2019}.
Specifically, MATLAB's stiff differential-algebraic equation solver \emph{ode15s} was used to be obtain numerical solutions to the system \cref{thinNS,thinGN,thinYL,thinBC,thinsym,thinArea}.
We added pseudo-time derivatives and pseudo-time coefficients, denoted by $\xi_1$, $\xi_2$, $\xi_3$, and $\xi_4$ to the system so that
the anchoring condition on the nematic--substrate interface \cref{thinNS},
the anchoring condition on the gas--nematic interface \cref{thinGN}, and
the nematic Young--Laplace equation \cref{thinYL}, take the forms
\begin{align}
\xi_1 \dv{\thetaNS}{t}&=K\left(\thetaGN - \thetaNS\right) + \CNS h \sin\thetaNS\cos\thetaNS, \label{numNS} \\
\xi_2 \dv{\thetaGN}{t}&=K\left(\thetaGN - \thetaNS\right) - \CGN h \sin\thetaGN\cos\thetaGN, \label{numGN} \\
\xi_3 \dv{h}{t}&=p_0 + h_{xx} + \frac{K}{2}\left(\frac{\thetaGN - \thetaNS}{h}\right)^2, \label{numYL}
\end{align}
respectively, and the area constraint \cref{thinArea} takes the form
\begin{equation}\label{numArea}
\xi_4 \dv{p_0}{t} = \dfrac{1}{2} - \int_0^1 h \, \dd x.
\end{equation}
The contact-line condition \cref{thinBC} and the symmetry and regularity condition \cref{thinsym} are unchanged.
We numerically solved the pseudo-time-dependent equations \cref{thinBC}, \cref{thinsym} and \cref{numNS}--\cref{numArea}, and then allowed the pseudo-time-dependent numerical solutions to approach a steady state, so that the solution of the system \cref{thinBC}, \cref{thinsym} and \cref{numNS}--\cref{numArea} approaches the solution of the system \cref{thinNS,thinGN,thinYL,thinBC,thinsym,thinArea}.

All of the numerical solutions presented in the present work used the initial condition $\thetaNS=\thetaGN=\pi(\sin(2\pi x)+1)/4$, $h=\hiso=3(1-x^2)/4$, and $p_0=\pI=3/4$, and pseudo-time coefficient values of $\xi_1=0.01$, $\xi_2=0.01$, $\xi_3=1$, and $\xi_4=0.01$.
Solutions for $\theta$ with a discontinuity at $x=\bar{x}$ ($0<\vert\bar{x}\vert<1$) can readily be generated by using the initial condition $\thetaNS=\thetaGN=\pi(\tanh(4(x-\bar{x})))/2$.
As described in \cref{sec:discontinuous},
numerical investigations of the system suggest that
(i) all solutions for the height of the ridge $h$ decrease monotonically from a maximum value at $x=0$, and have even symmetry about $x=0$,
(ii) all continuous solutions for the director angle $\theta$ also have even symmetry about $x=0$, and
(iii) solutions for $\theta$ with a discontinuity only at $x=0$ have odd symmetry about $x=0$.



\begin{thebibliography}{10}
\bibitem{Notei}
We note that a closed-form parametric solution to the system \cref{thinNS,thinGN,thinYL,thinBC,thinsym,thinArea}
can be obtained (see Cousins \cite{Cousinsthesis} for details),
but is not particularly useful in practice and so is not described here.
\end{thebibliography}

\end{document}